# High ENSO-based 18-month lead Potential Predictability of Indian Summer Monsoon Rainfall

Devabrat Sharma[1,3], Santu Das[1,3] and B. N. Goswami[2*]

[1]Institute of Advanced Study in Science and Technology (IASST), Guwahati, India
[2]Department of Physics, Cotton University, Guwahati, India
[3]Academy of Scientific and Innovative Research (AcSIR), DST – IASST, Guwahati, India

*Corresponding Author

*Scientific basis for long-lead seasonal prediction of Indian summer monsoon rainfall (ISMR) critical for water resource and crop strategy planning is lacking. Using a new predictor discovery method, here we show that the depth of $20^0$-isotherm (D20) is least influenced by atmospheric 'noise' and that the 18-month lead forecasts of ISMR have high potential skill (r = 0.86). The high potential predictability is due to smaller initial errors associated with the 18-month lead initial conditions and their 'slow' growth associated with the El Nino and Southern Oscillation (ENSO). The potential skill arises not only from the correlation between ISMR and large-scale slowly varying D20 but also contributed significantly by that with the interannual small-scale D20 anomalies indicating a seminal role of the nonlinearity on the potential predictability. It is, therefore, imperative that a nonlinear predictor discovery as well as nonlinear prediction model is essential for realizing this potential predictability.*

The total amount of rainfall during June-September over the continental India representing the intensity of the Indian summer monsoon rainfall (ISMR) and accounting for more than 75% of annual rainfall over the country is the primary water resource for agricultural productivity in the region. As a result, large-scale 'floods' and 'droughts' associated with modest year-to-year variations of ISMR exceeding 10% of the long-term mean cause extreme socio-economic distress for the people of the region[1–3]. Prediction of ISMR is, therefore, awaited with bated breath by policy makers and farmers alike and influence market sentiments[4]. It is, therefore, not surprising that improvement of skill of Indian monsoon weather and climate translates to direct economic benefit to farmers and fishermen[5]. As a result, past large swings of ISMR lasting for longer periods (~a century) have been known to be associated with flourishing or perishing society in the region[6]. Forewarning about ISMR one year or longer lead times is, therefore, critical for water resource and alternative cropping strategy planning and food security of the region.

The scientific basis for ISMR predictability has been founded on a strong footing[7] as a sub-set of seasonal predictability of tropical climate in general[8,9] being modulated by predictable drivers like the El Nino and Southern Oscillation (ENSO) and/or Atlantic Multi-decadal Oscillation (AMO), Pacific Decadal Oscillation (PDO) etc. While significant advances have been made in understanding the physics of the Indian summer monsoon, its variability and predictability[2,7,10–18] over the past century, the prediction of seasonal mean ISMR even one month in advance remained a grand challenge problem[19]. Recognizing the importance and urgency for such forecasts, Blanford developed the first empirical forecast model for ISMR in 1884[11]. Since the time of Blanford, even with considerable research to improve the skill of empirical forecasts of ISMR and attempts to operationalize them, the skill of such predictions remained poor and far below limit on potential predictability[8,20–31]. For example, the correlation skill of the official operational forecasts made by India Meteorology Department (IMD) empirical models for 1989–2012 is -0.12[30]. Examining hind cast skill of JJA precipitation over the Indian monsoon region from 8 different seasonal prediction systems based on about 30 years of retrospective forecasts under the Climate Historical Forecast Project (CHFP), Jain et al. (2019)[32] note that the skill of few dynamical prediction systems is as high as 0.6 while that of most prediction systems are much lower. As a result, the multi-model ensemble (MME) skill also remains close to 0.6. In their study Saha et al. (2019)[33] argue that the potential correlation skill for ISMR is ~0.82, much higher than earlier estimate of ~0.65 using signal to noise ratio or 'perfect model' experiments[34]. Thus, the skills of all empirical or dynamical models one season in advance remain significantly lower than the potential predictability. More importantly, attempts to predict one year or longer in advance has been lacking. Further, there has been no study indicating the potential limit on predictability at long leads, e.g. 9-month lead forecasts, 12-month-lead forecasts or 18-month lead forecasts. Here, we present a new predictor discovery method and unravel that the ISMR has high potential predictability even at 18-month lead.

The predictability of ISMR emerges primarily from its strong association with the slowly varying coupled ocean-atmosphere processes like a research-discharge oscillator[35] linked with the ENSO, the Atlantic Nino[36] and the Indian Ocean Dipole mode[37–39] in the tropics. The contributions from the tropical predictors, however, can be modulated by multi-decadal ENSO[40], the AMO and the PDO. As indicated by a large number of previous studies, the potential predictability of ISMR arising from association with the slowly varying predictors is high but all prediction systems (empirical as well as dynamical) failed to realize it. The limiting factor for predictability arises from 'internal' variability of the coupled system appearing as 'noise' embedded in the predictors. The 'noise' originates from the nonlinearity of the slowly varying predictors (event-to-event variability, asymmetry between positive and negative phases, contributions to the large-scale slow oscillatory component from small-scale fast processes etc.). The 'linear' contribution of the predictors (indicated by linear correlation between predictor and ISMR) is limited to explaining only about 40% of ISMR variability from ENSO[41]. Recognizing these nonlinear models using deep learning and artificial intelligence are emerging for ISMR prediction[42–44]. However, the field is still in its infancy. For the prediction systems based on dynamical models, the 'nonlinearity' is already built in. However, the skill of predictions compared to observations is still limited due to the biases in simulating the observed climate by the models. As these biases of the coupled modes are reduced, the skill of seasonal predictions of ISMR improves[33]. Although our new predictor discovery method is essentially linear, it is able to estimate the potential limit on predictability. In addition, it provides insight into why a 'linear' method for prediction fails to come close to realizing the potential limit of predictability.

The philosophy of our predictor discovery method has been to identify a field that is least influenced by atmospheric 'noise' and find a method to maximize the contributions to the predictable signal from all three-ocean basins. The maximum simultaneous correlations with predictors based on sea surface temperature (SST), mean sea level pressure (MSLP) etc. does not reach higher than 0.65 due to intrinsic atmospheric 'noise'. We find that the thermocline anomaly as measured by the depth of $20^O$ isotherm (D20) is least influenced by atmospheric 'noise' and a predictor based on global tropical D20 of December (-2) at an 18-month lead provides a skill of 0.86 based on a 88 partially independent hindcasts. The ability of the predictor to estimate the potential limit on predictability appears to be due to the fact that the D20 for Dec (-2) fluctuate with the ISMR coherently over the entire tropical belt thereby maximizing the contribution to ISMR predictability from all three basins. The success of the 18-month lead forecasts appears also to be due to the fact that the Dec (-2) D20 seems to contain a state of the recharge-discharge oscillator that evolve to surface manifestation of ENSO (Atlantic Nino and IOD) after 18 months at May (0) to influence ISMR through simultaneous atmospheric bridge. The nonlinearity of the recharge-discharge oscillator would lead to growth of small errors in the initial state (Dec (-2) D20) and contaminate the initial condition at May (0) making the short-lead forecasts less skillful.

**A New Predictor Discovery Method**

Precursors for predicting ISMR (predictors) have generally been identified through associations found through linear correlations. Most important amongst them has been the association with the El Nino and Southern Oscillations (ENSO)[45–47] and remained the single most dominant predictor for ISMR over the past century. In the recent decades, a couple of other significant tropical associations, one with the Atlantic Nino[36,48–50] and another with the Indian Ocean Dipole (IOD)[37,39] have been identified. The predictability of ISMR is expected from the associations with these slowly varying tropical climate modes driven by coupled ocean atmosphere interactions. In addition to these tropical sources of predictability, significant associations of ISMR with extra-tropical sea surface temperature (SST) over the north Pacific associated with the Pacific Decadal Oscillations (PDO)[51,52] and that over the north Atlantic associated with Atlantic Multi-decadal Oscillations (AMO)[53–55] have emerged in the recent years.

Traditionally, a predictor is identified by using correlation maps between ISMR and relevant variables such as the SST and creating an index of the predictor by averaging over a region of largest significant correlations (e.g. the Nino3.4 for the ENSO). However, such a predictor would fail to represent simultaneous contributions of other predictors (e.g. IOD or Atlantic Nino). At any given time, the contributions from other predictors could also either be additive or could interfere with each other. In order to maximize the contributions all three potential tropical predictors (the ENSO, IOD and Atlantic Nino), we use the full global correlation map and create a predictor time series by projecting the anomaly (e.g. SST, HC or D20) on the significant part of the correlation. Our predictand being JJAS mean ISMR, we would like to identify predictors up to 24-month lead predictions. In this manner, we create three predictors namely, $D_p$ (i,j), $H_p$ (i,j), $S_p$ (i,j), based on projections of the three fields namely monthly means of D20 anomaly, upper ocean heat content anomaly (HC) and SST anomaly, where i = 1 to 24 refers to lead months while j = 1 to 138 refers to years from 1871 to 2008, 1872 to 2009 and 1873 to 2010. (See Methods for details).

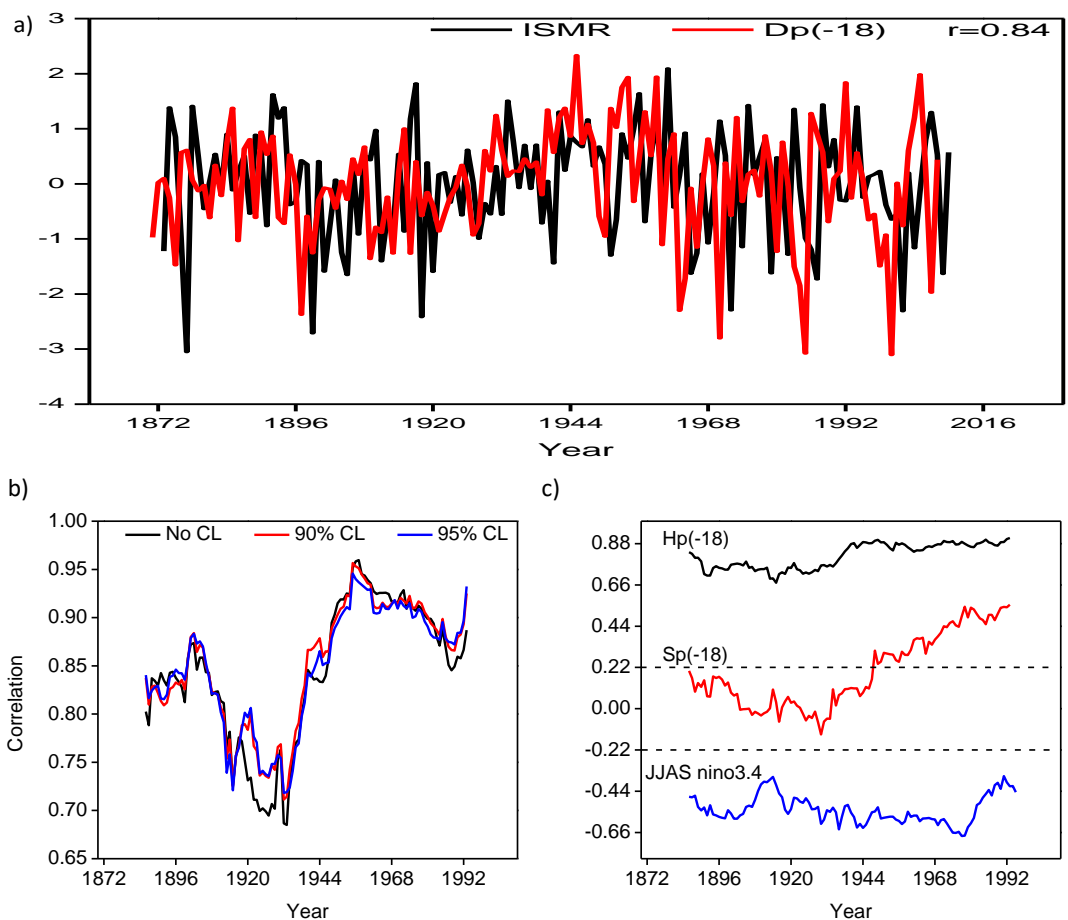

*Figure 1: (a) Time series of normalized ISMR anomaly and $D_p(-18)$ i.e the D20 monthly anomaly based predictor with 18-month lead corresponding to Dec (-2) for the period 1871 to 2008. Simultaneous correlation between the two indicated. (b) 31-year moving correlation between ISMR and $D_p(-18)$ constructed using projections on the full tropical correlation pattern (no CL), on pattern exceeding 90% CL and on pattern exceeding 95% CL. (c) 31- year moving correlations between ISMR and $H_p(-18)$ and with $S_p(-18)$ (dashed lines represent statistically significant correlation).*

A major finding of our exercise is that correlations between $D_p$ and ISMR have a maximum at 18-month lag corresponding to Dec (-2) initial conditions with r=0.84 (Figure 1a). This unprecedented high correlation for any predictor over a period of 138 years is highly significant as the correlations between ISMR and D20 anomaly at this lag at any point over the domain (see Extended Data Figure 1) is much smaller than that between the predictor, $D_p$ and ISMR validating the basis for our predictor discovery method that the projections maximizes contributions from all three basins to ISMR predictability. It is notable that the correlations between $D_p$ at 18-month lag (Dec (-2) initial condition) and ISMR with $D_p$ based on projections made on the full correlation map (no CL) are nearly identical to those made based on correlations exceeding 90% or 95% CL (Figure 1b).

As the ENSO and ISMR relationship is known to undergo epochal variations[41], we examine possible epochal variation in the ISMR and D20 predictor ($D_p$) with 31-year moving correlations between ISMR and $D_p$ at 18-month lead (Figure.1c). We contrast the same with similar 31-year moving correlations between ISMR and $S_p$ at 18-month lead (Figure.1b). It is also contrasted with simultaneous moving correlations between ISMR and JJAS mean Nino3.4 SST (Figure 1c). It is interesting to note that the epochal variation of the correlations between ISMR and D20 predictor is smaller in amplitude than that associated with correlations between ISMR and SST predictor ($S_p$). It may be noted that most empirical models including the operational model of India Meteorology Department has been poor during the recent years between 1989-2012 and has been attributed to the inability of the models to represent the teleconnection with the CP El Nino[30]. In this backdrop, it is notable that the correlation between ISMR and the D20 predictor remains at a level of ~0.9 during the past five decades. This finding has two important implications. Firstly, we expect potential hind cast skill of more than 0.8 at the lead of 18-months. Secondly, our study indicates that the ENSO-ISMR relationship remains robust in recent decades in contrast to studies using SST to represent ENSO and ISMR relationship concluding that ENSO-ISMR relationship weakened significantly during recent decades[56]. It is also notable that in contrast to statistically significant negative correlations between ISMR and JJAS Nino3.4 SST, the correlation between ISMR and $S_p$ at 18-

month lead is insignificant during the earlier period up to ~1960 while becoming increasingly positive and significant during recent decades. The increasing positive correlation between global SST based predictor $S_p$ and ISMR may be due to the increasing atmospheric moisture with global warming.

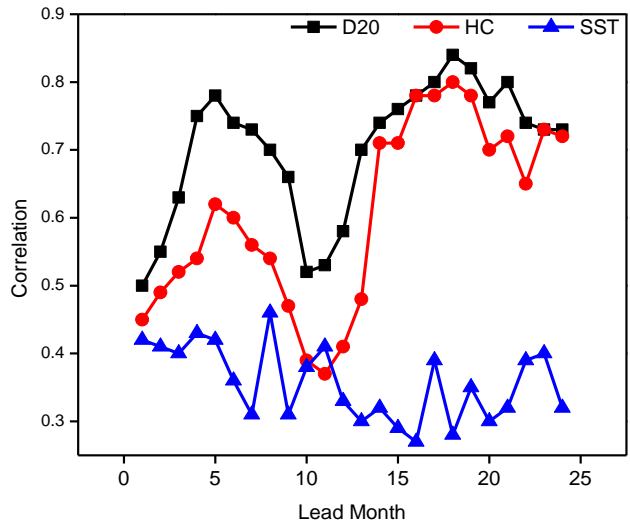

*Figure 2: Simultaneous correlations between ISMR and $D_p$, $H_p$, $S_p$ for all leads.*

The correlations between $D_p$ and ISMR at all 24 leads indicate that there are two maxima, one at 5-month lead and another at 18-month lead (Figure 2). The correlations between ISMR and $H_p$ has same variations with lead time as those with $D_p$ but are always smaller than those with $D_p$. Similar correlations with $S_p$, however, are not only much smaller than either $D_p$ or $S_p$, they monotonically decrease with lead time. The findings suggest that the $D_p$ is least affected while $S_p$ is most affected by atmospheric climate 'noise'. While the $H_p$ is more affected than the $D_p$, it is much less affected than the $S_p$. Although $S_p$ is constructed in exactly the same way as $D_p$ by projecting on the global correlation maps, the significant differences in correlations with the ISMR indicate that our findings of high correlations between ISMR and $D_p$ are non-trivial.

## Long-Lead Potential Forecasts skill

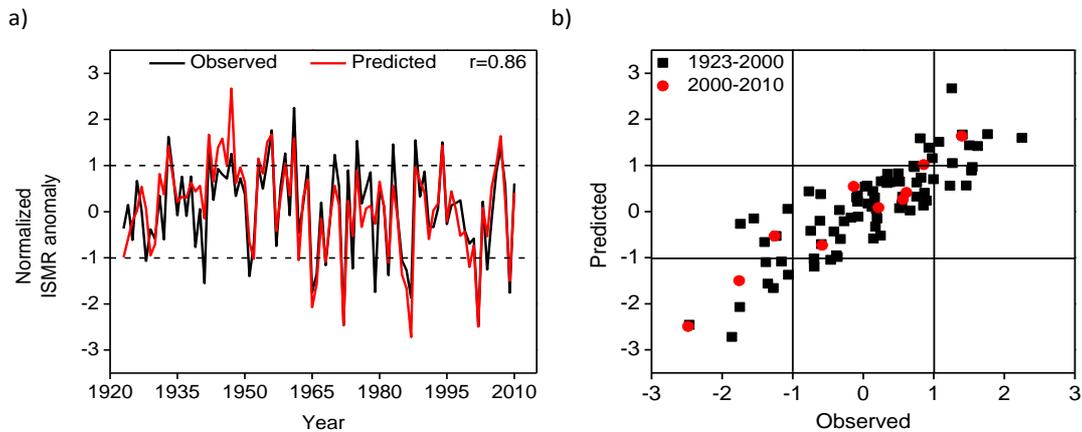

*Figure 3: Verification of hindcasts. (a) Predicted normalized ISMR anomaly for 18-month lead 88 year hindcasts together with corresponding observed ISMR anomaly. (b) Scatter plot of predicted and observed normalized ISMR anomalies for 18-month lead hindcasts.*

Armed with the correlations between ISMR and D20 predictors with various lags, we construct a simple linear regression model for predicting ISMR at 1-month lead using the D20 predictor at 1-month lag namely for the May (0) and two similar models for predicting ISMR at 5-month lead and 18-month lead using D20 predictors for Jan (0) and Dec (-2). Following a technique similar to Wang et al. (2015)[30], we use the first 50-year period (1871-1920) for training the models and make the hindcasts for the next 5 years. Then move the training window forward by 5 years and make hindcasts for the next 5 years and so on. In this manner, an ensemble of 88 years of retrospective forecasts are made and validated with observations for the 18-month lead hindcasts (Figure.3). Similar verifications 1-month lead and 5-month lead hindcasts may be found in Extended Data Figure 6) and summary of performance of the three models is presented in Extended Data Table 1. We note that the 18-month lead hindcasts are the most skillful with a skill of R=0.86 and RMSE= 41.94mm. In contrast to tendency of most empirical models to underestimate the interanual variance[57] missing most of the extreme events, this model predicts the interannual standard deviation of ISMR close to observed (80.31 mm compared to observed 77.62 mm). As a consequence, almost all real extreme droughts and floods (>1.5 s.d and <-1.5 s,d) are well predicted. However, two moderate floods are underestimated while five moderate droughts are underestimated. It is also notable that barring two cases, no 'normal' monsoon is predicted as 'flood' (false alarm) while only one 'normal' is marginally predicted as 'drought'. Overall, the skill of 18-month lead hindcasts (Figure.2) is quite impressive as it is probably close to the limit of potential predictability of ISMR[33]. The skill of 5-month lead hindcasts (r=0.81, rmse=47.20 mm, Extended Data Figure 6 b, d) are also impressive as it is better than any known dynamical or empirical forecast systems at this lead evaluated with such large sample of partly independent hindcasts (88-years). The interannual standard deviation of predicted ISMR is also close to observed (75.08 mm). While one 'flood' and 6 'droughts' are underestimated in this case, there are four false 'flood' alarms' and three false 'drought' alarms. The skill of 1-month hindcasts (r=0.66) while being exactly in the ballpark of best of current operational predictions systems[32,58], is significantly poorer than the 5-month and 18-month lead hindcasts in all counts such as underestimation and false alarms of 'droughts' and 'floods'. While counter intuitive, the much higher skill of the 18-month lead model compared to those of 5-month and 1-month lead models is intriguing and may be the most important finding of this study.

However, the predictors are based on the correlation map constructed with the full data. Although the training period is independent of the forecast period, predictor finding is not. Therefore, these hindcasts are 'partially independent' hindcasts. As indicated by Delsol and Shukla (2009)[59] artificial skill may be built in the forecasts if the full data is involved in the predictor discovery. Our objective in this study is to examine the potentially achievable skill and not build a prediction model and argue that these 'partially independent hindcasts' gives us an estimate of 'potential skill'. The correlation between ISMR (1873-2010) and $D_p$ (-18) (1871-2008) (r = 0.84) also provides us a rough estimate of the 'potential skill'. This estimate of potential skill is independent of the predictor discovery method as the correlation between ISMR and $D_p$ constructed from a correlation map between ISMR (1930-1979) and $D_p$(-18) (1928-1977) also gives a high potential skill of 0.88.

**Recharge-Discharge Oscillator and basis for long-lead predictability**

The growth of small errors associated with ENSO forecasts are dependent on the initial conditions in relation to the evolutionary cycle of the ENSO known as the 'spring predictability barrier'[60,61]. Therefore, growth of errors for ISMR predictions based on ENSO recharge discharge oscillators would also be dependent on initial conditions associated with different phases of evolution of the oscillator. In order to estimate the initial error associated with the initial conditions associated with the 18-month lead predictions, we try to identify the phase of the recharge discharge oscillator with which this initial condition is related.

The dominant EOF of the NDJ mean D20 anomalies over the tropical basin for the period 1871 – 2010 (Extended Data Figure 7a) represents the eastern Pacific (EP) El Nino with dipole pattern of depression (elevation) of thermocline in the eastern (western) Pacific. It is notable that it is associated with an IOD type pattern with elevation (depression) of thermocline in the eastern (western) Indian Ocean. The fact that the dominant EOF represents interannual variations of warm water volume associated with the ENSO is supported by a strong correlation between the PC1 (Extended Data Figure 7d) and NDJ Nino3.4 SST anomalies (r =0.88). The EOF2 (Extended Data Figure 7b) on the other hand represents a strong multi-decadal modulation of the central Pacific (CP) El Nino with an approximate period of approximately 80 years (Extended Data Figure 7c). It is also notable that the dominantly interannual variations of PC1 are weakly modulated by a multi-decadal oscillation with approximately 80-years. ENSO is known to have a multi-decadal component[40] and modulation of PC1 and the PC2 may be related to this component of the ENSO variability. To test this possibility, we define a ENSO multi-decadal (ENMD) index as,

ENMD (y) = NDJ SST anomaly averaged between (150W-90W, 20S-20N) – NDJ SST anomaly averaged between (130E-150W, 20N 55N) with year (y) going from 1871 to 2010.

The normalized 11-year moving average of ENMD compared to normalized PC1 (Extended Data Figure 7e) indicates that indeed it appears that the major swings of PC1 are related to major swings of the ENMD. A similar EOF analysis of NDJ HC anomalies (Extended Data Figure 8) indicates a close link between both the EOFs of D20 and HC with correlation between PC1 of D20 and that of HC being r = 0.99.

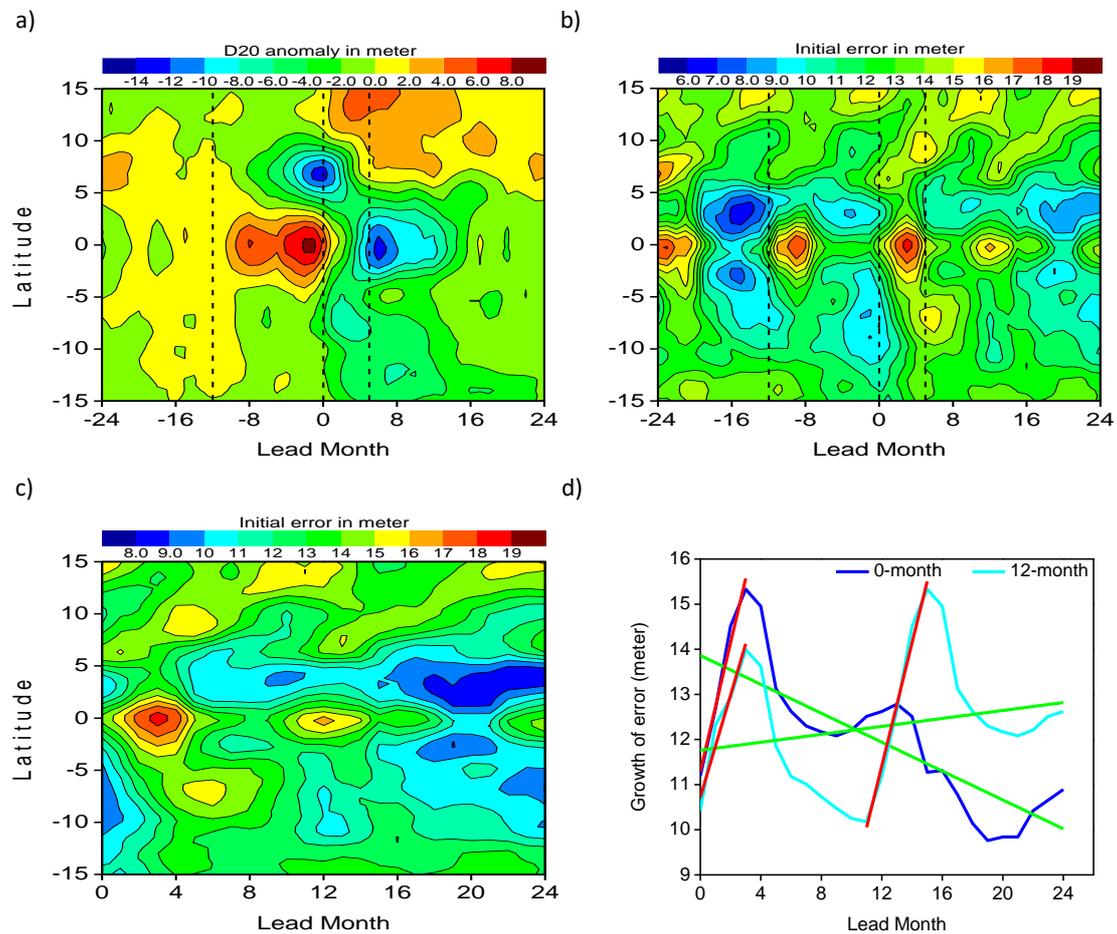

*Figure 4: (a) Zonal mean over the Pacific basin (150E -90W) of lead-lag composites of monthly D20 anomaly (m) with respect to peak ENSO events identified from PC1 of NDJ D20 anomaly (Extended Data Figure 7c) as a function of lead-lag and latitudes representing the evolutionary cycle of the recharge-discharge oscillator. (b) Zonal mean over the Pacific basin of standard deviation of composites of monthly D20 anomaly (m) with respect to peak ENSO events identified from PC1 of NDJ D20 anomaly based on 20 events. The standard deviation or the event-to-event variability of initial conditions corresponding to each forecast is a measure of 'initial error' corresponding to that warm water volume initial condition. (c) Zonal mean over Pacific basin of estimated growth of initial error (standard deviation of divergence of trajectories) up to a lead of 24-months for an initial condition corresponding to peak of El Nino event (Dec (-1)) as a function of latitudes. (d) Zonal mean and latitudinal mean between 7⁰S and 5⁰N of estimated growth of errors for an initial condition corresponding to peak of El Nino event (Dec (-1)) and for another initial condition corresponding to 12-months before the peak of El Nino event (Dec (-2)). A fast rate of growth (red) and slow rate of growth (green) of errors are indicated.*

There are 20 strong positive ENSO events with normalized PC amplitude larger than +1 (Extended Data Figure 7d). Lead-lag composites with respect to the peak of the events Dec (-1) w.r.t ISMR of following year (0-year) are constructed going backward up to 24 months from the peak and 24 months forward from the peaks. The composites (Extended Data Figure 9) show similar eastward propagation of the warm water volume as in the correlations between ISMR and D20 (Extended Data Figure 1). The zonal mean D20 anomaly (Fig.4a) shows the evolution of the recharge-discharge oscillator with asymmetry between the positive (El Nino) and negative (La Nina) phases. As lead-lag in Figure 4 is with respect to the peak month of ENSO event (Dec (-1)), lag=0 and lag =-12 correspond to initial conditions for 5-month and 18-month lead forecasts for ISMR respectively. Also the rate of transition from the positive peak to a negative peak is much faster than that from the negative peak to a positive peak. It is notable that the maxima in potential skill for 5 and 18-month lead forecasts (Fig.1d) are close but not at the minima of initial errors (event-to-event variability) (Fig.4b). This appears to be due to the fact that the growth of errors associated with the initial conditions associated with the minima of initial errors (September/August) is larger than those associated with 5-month and 18-month lead forecasts (December/January). In addition to the initial error, the skills of forecasts also depend on the rate of growth of errors associated with the initial conditions. To estimate the growth rate of errors corresponding to these initial conditions, the divergence of trajectories of D20 up to 24-month lead initiated with same initial condition but with different initial error (coming from 20 different events) is estimated use a methodology described in Goswami and Xavier (2003, see Methods for detail)[62].

Zonal mean of growth of errors as a function of latitude (Fig.4c) indicate that within the equatorial belt between 7°S and 5°N, while the errors grow rapidly for first six months, it oscillate on an annual time scale with an overall decreasing trend with increasing lead months. In the discharge phase of the oscillator, however, the errors seem to grow while propagating away from tropics on both hemispheres at the rate of about 0.85°/month. Similar growths of errors are estimated for three other initial conditions corresponding to 3-month before, 12-month before and 16-month before the peak ENSO events (Extended Data Figure

10). The growth of average error over the equatorial domain (7°S and 5°N) for the initial condition corresponding to the peak ENSO event indicates convergence of trajectories with increasing forecast lead consistent with decaying phase of the oscillator (Figure 4d). The growth of errors corresponding to an initial condition 12-months before the peak, however, shows divergence of the trajectories as expected in a growing phase of the oscillator. It is also noted that consistent with earlier studies[60,63], the growth rate of the errors in the coupled system is governed by two time scales, one 'fast' time scale (red in Fig.4d) associated with the coupled instability while another 'slow' time scale (green) associated with the coupled slow oscillator. It is also notable that while the initial error associated with the initial condition corresponding to 12-month before the peak is higher than that associated with the initial condition corresponding to 16-month before the peak, the rate of error growth in the long time scale is slower for the one corresponding to 12-month before the peak compared to that for the one 16-month before the peak (Extended Data Figure 11). It may be noted that all correlations between ISMR and $D_p$ for 18-month lead forecasts to 24-month forecasts are high and comparable (Fig.2) and consistent with low initial errors (Fig.4b). The differences in correlations between ISMR and $D_p$ for these lead forecasts are also consistent with magnitude of initial errors and rate of their slow error growth. We recognize that our initial error estimate may have some 'uncertainty' from observational errors. The differences in actual magnitude of error for 5-month, -18-month and 22-month forecasts could partly be attributed to the uncertainty in the initial errors.

# Challenges in realizing the Potential Predictability

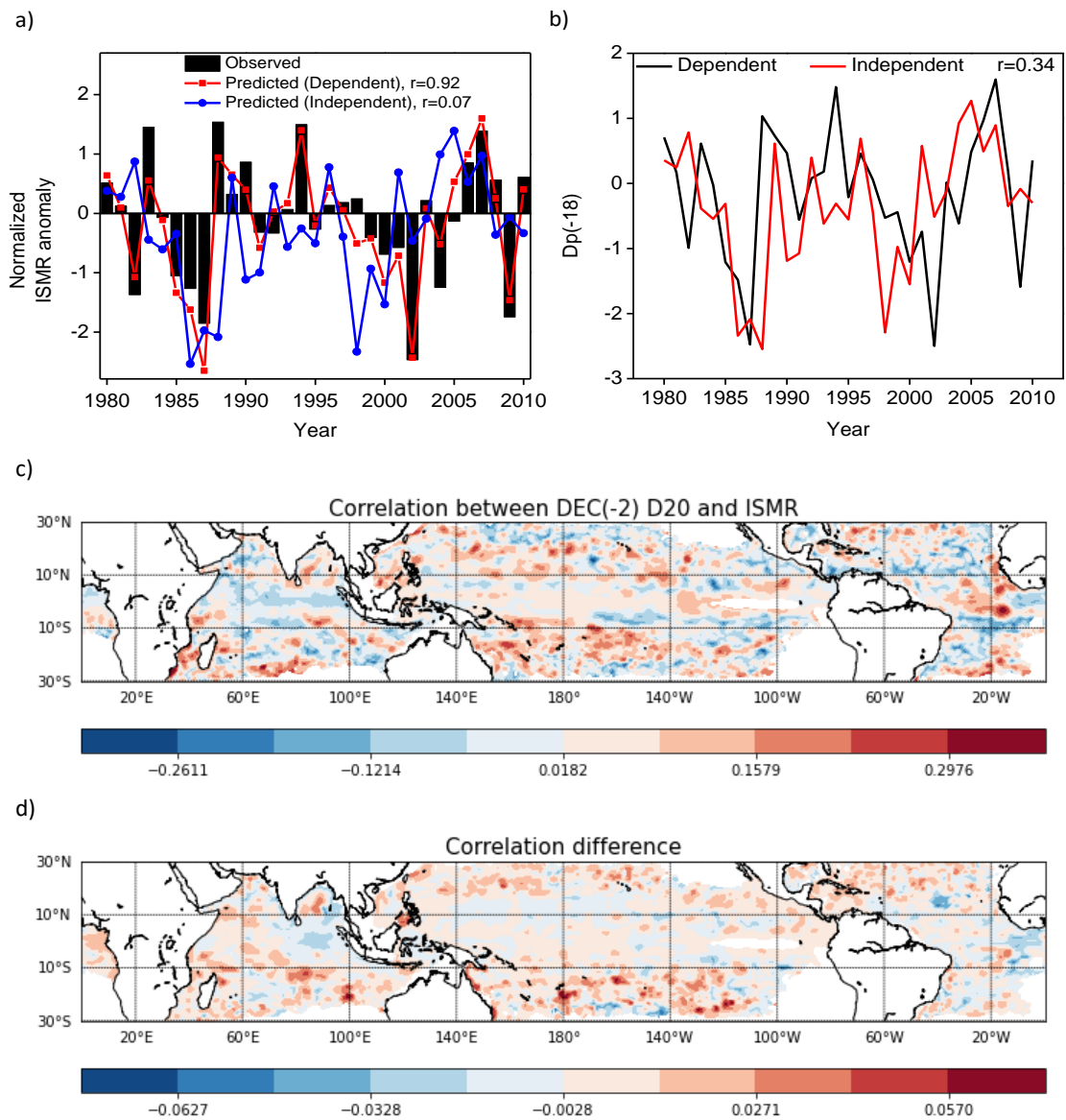

*Figure 5: (a) Observed (black) and predicted (red) normalized ISMR anomaly between 1980 and 2010 when the construction of D20 predictor included the forecast period showing strong correlation. Also shown is when the forecast period was kept completely independent (blue) showing poor correlation with observed ISMR. (b) D20 predictor (projection on the correlation maps) when the construction of D20 predictor included the forecast period (black) and when the forecast period was kept completely independent (red). (c) Correlation between D20 anomaly (December (-2)) and ISMR over the period 1871 and 1980. (d) Difference between two correlations maps, one shown in (c) calculated with data between 1871 and 1980 and another calculated with data between 1871 and 1981.*

While the potential predictability of the 18-month lead D20 prediction is high, it is found that it would be extremely difficult to realize the same in real time forecasts. Keeping the period 1980-2010 of ISMR out of the predictor discovery, fully independent forecasts are made for the period 1980 and 2010. For this purpose, D20 predictors are found based on map of correlation between D20 (Dec (-2)) and ISMR for the period between 1871-1977 and 1873-1979 respectively, a regression model is then trained for the immediately past 50-years (1929-1977) to predict for ISMR (1980) using 1978 DEC (-2) D20. The correlation map is updated with 1978 DEC (-2) D20 and 1980 ISMR data and D20 predictor (Dec (-2)) is calculated by projecting on the new correlation map. The model is trained with past 50-year of updated data and prediction is made for ISMR (1981) and so on. The skill of prediction of fully independent data comes out to be far short of the potential predictability (Fig. 5a, r = 0.07). The poor skill of independent predictions is consistent with large differences between the 'perfect' initial conditions and 'real' initial conditions (Fig.5b, r = 0.34). The D20 predictor is based on the projection of D20 anomaly on the correlation map between D20 and ISMR. Such a map for the period 1871-1980 (Fig.5c) shows that embedded in the large-scale ENSO pattern has some small scale correlations arising from small scale variability of D20. The large-scale smooth pattern tends to have an eastward propagation (Extended Data Figure 1) while propagation of the small-scale patterns is unclear. As a result, the difference between the correlation maps for the period 1871-1980 and that between 1871-1981 (Fig.5d) shows that while the large-scale ENSO pattern may not change significantly in one year, the small-scale correlation pattern changes significantly. It is also seen that the differences in correlations are about 5 times smaller than the full correlations with maximum differences being in the off-equatorial sub-tropics. It may, therefore, be concluded that even after minimizing the atmospheric 'noise', the 'nonlinearity' in the D20 is critical in achieving potential prediction skill associated with the ENSO-ISMR relationship. A corollary to it is that a linear model may never be able to attain the potential predictability. Therefore, a 'nonlinear' model is a necessary condition for achieving the potential predictability.

**Discussion**

Conventional wisdom of predictability based on growth of initial errors with dynamical prediction systems indicates that one should expect decreasing skill of forecasts with increasing lead times. In the context of ISMR prediction associated with ENSO-ISMR relationship, it would depend on the two time scales of growth of errors for ENSO prediction[60]. On shorter time scales (1-6 months), the error grows due to the fast time-scale of growth associated with the instability while on a slower time scale errors decay in the decaying phase of the slow oscillation while it builds up in the growing phase of the slow oscillation. As the typical period of the ENSO is ~ 4 years, these decaying or growing phases of error growth could last for about 2-years. Depending on the initial error and growth rate of the 'slow growing' error, there is potential in predicting skillful ENSO forecasts and ISMR forecasts up to 24-months in advance.

One potential advantage of empirical models may be that if the predictor has highest correlation with the predictand at long leads, we may actually expect poorer skill at short leads due to being in the 'fast error-growing phase' and still could have higher skill at long lead due to being in the 'slow error-growing phase'. Our predictor discovery methodology and the selection of D20 as the predictor have been successful in identifying the lag at which the coupled recharge-discharge oscillator kicks in the highest response to the ISMR and find that the 18-month lead forecasts have potentially the highest skill (r =0.86) consistent with initial error and error growth with the particular initial condition. However, realizing the potential predictability has a major roadblock. We find that the potential high skill arises not only from the correlation between ISMR and large-scale slowly varying D20 but also contributed significantly by that with the small-scale fast changing D20 anomalies indicating seminal role of the event-to-event variability and nonlinearity on the potential long lead predictability. Therefore, it is imperative that a nonlinear predictor discovery as well as a nonlinear prediction model is essential for realizing this potential predictability. While the dynamical models are following this path, their skill is still limited by their biases in simulating the climate. So far, nobody has attempted and tested the efficacy of 18-month lead ISMR forecasts using a state of the art coupled modeling system. Our finding suggests that it should be explored. With the reduction of biases in the coupled models in the coming years, it may be possible to realize the potential of the 18-month lead ISMR forecasts by dynamical models. For empirical prediction, our findings highlight the need for using nonlinear techniques like artificial intelligence (AI) and deep learning to train the models to patterns associated with the asymmetry and event-to-event variability of the ENSO. Success of such techniques in long-lead ENSO prediction[64] is suggestive of potentially similar success in long-lead ISMR prediction using such techniques.

Acknowledgements:
BNG is grateful to the Science and Engineering Reassert Board (SERB), Government of India for the SERB Distinguished Fellowship. DS acknowledges IASST for the Junior Research Fellowship and AcSIR for the PhD program. SD acknowledge AcSIR for the recognition and IASST for providing support to carry out the research work.

**Methods**

*Data*

Indian summer monsoon rainfall (ISMR) is defined as the June-September accumulated rainfall over land points of continental India constructed from monthly mean rainfall data[65] based on a fixed network of 306 stations. The data is available between 1871 and 2010. Anomaly of ISMR is defined as departure from long-term mean and standardized by its own interannual standard deviation. Global gridded monthly mean sea surface temperatures are obtained from monthly SST analysis; the Centennial in situ Observation-Based Estimate of SSTs (COBE SST2)[66] for the same period (1871-2010). Amongst several century scale SST analysis data sets based on World Ocean Database (WOD) using slightly different analysis techniques, COBE-SST2 has been found to be closest to observations[67]. Monthly anomalies are calculated as departures from long-term climatological annual cycle at each grid point. Depth of the $20^0$ isotherm (D20) and heat content of upper 200 meters (HC) over the global tropics (0 -360$^0$E, 30$^0$N – 30$^0$S) are obtained from Simple Ocean Data Analysis version 2.2.4 (SODA-2.2.4)[68,69]. Version 2.2.4 represents their first assimilation run of over 100 years and uses the 20CRv2 winds. The ocean model is based on Parallel Ocean Program physics with an average 0.25°x0.4°x40-level resolution. Observations include virtually all available hydrographic profile data, as well as ocean station data, moored temperature and salinity time series, surface temperature and salinity observations of various types, and nighttime infrared satellite SST data. The output is in monthly-averaged form, mapped onto a uniform 0.5°x0.5°x40-level grid and available for the same period (1871-2010). Monthly anomalies are calculated in a similar manner as departure from long-term climatological annual cycle. Linear trends are removed from all data.

*Predictor discovery Algorithm*

In order to maximize the contribution to predictability of ISMR from all the three basins and to minimize the contamination of atmospheric 'noise', let us isolate predictors based on D20, heat content (HC) and SST. Let

D(i, j) = D20 anomaly over tropical belt (30S-30N, 0-360) for $i^{th}$ lag at 1, 2, ... 24 month lags (w.r.t ISMR) representing the D20 for MAY (0), APR (0), MAR(0), .... DEC(-1) ...... JAN(-1), DEC(-2), ..... SEP(-2), .. JUN (-2) where 0 represents the year of prediction while -1 and -2 represents one or two years prior to the year of prediction. Subscript 'j' refers to the years from 1871 to 2008, 1872 to 2009 and 1873 to 2010 (1,... ,138).

H(i,j), and SST (i,j), represent monthly heat content and SST anomalies up to 24-months lag with respect to the ISMR.

CorD (i) = Correlation maps between ISMR and monthly D20 anomalies at all lags (Extended Data Fig.1) indicate an eastward propagation of a negative thermocline anomaly signal during the first 12-months while a positive thermocline anomaly signal during the next 12-months in the tropical Pacific associated with the recharge-discharge oscillator. An empirical orthogonal function (EOF) analysis with all lags (Extended Data Fig. 2a) brings out notable insight on the role of the recharge-discharge oscillator and its teleconnection with the ISMR. The first of the two dominant patterns of correlation (EOF1, Extended Data Fig.2a) represents the teleconnection with the EP El Nino with a quasi-biennial year period while the second pattern (EOF2, Extended Data Fig.2b) represents a multi-decadal modulation of ENSO. It is noteworthy that the EOF1 embodies teleconnection with the IOD as well while the EOF2 embodies similar teleconnection with the Atlantic Nino.

CorH(i) for i=1,2,...,24, are the maps of correlation between ISMR and HC anomaly (Extended Data Fig.3). As expected the CorH patterns are very similar to those of CorD and similar eastward propagation of heat content associated with the thermocline wave is seen. The dominant EOFs (Extended Data Fig.4) are also similar to those of CorD. Similar maps for correlation between ISMR and monthly SST anomaly at lags up to 24-months (CorS(i)) are also calculated (not shown).

Next, the monthly anomaly field is projected on the correlation pattern significantly at greater than 95% confidence level over the whole domain for all lags. Let,

$D_p(i,j)$ => Projection of monthly D20 anomaly for the $i$th lag and $j$th year with i = 1,2, ....24 and j= 1, 2, ....., 138.

$H_p(i,j)$ => Projection of monthly HC anomaly for the $i$th lag and $j$th year with i = 1,2, ....24 and j= 1, 2, ....., 138.

$S_p(i,j)$ => Projection of monthly SST anomaly for the $i$th lag and $j$th year with i = 1,2, ....24 and j= 1, 2, ....., 138.

We find that correlations between $D_p$ and ISMR have two maxima, one at 5-month lag corresponding to Jan (0) initial conditions with r=0.78 (Extended Data Figure 5a) and another at 18-month lag corresponding to Dec (-2) initial conditions with r=0.84. It is interesting to note that the correlations between ISMR and D20 anomaly at this lag at any point over the domain (see Extended Data Figure 1) is much smaller than that between the predictor, $D_p$ and ISMR validating the basis for our predictor discovery method that the projections maximize contributions from all three basins to ISMR predictability. We find that unlike $D_p$ at 18-month lag Dec (-2) initial condition, the correlations between ISMR and Dp based on projections made on the full correlation map (no CL) are significantly lower than those made on correlations exceeding 90% or 95% CL (Extended Data Figure 5b). It is also noteworthy that although the correlation patterns between D20 anomaly and HC anomaly at all lags are nearly identical (Extended Data Figures 1 and 3), the correlations between ISMR and HC based predictor ($H_p$) are slightly lower than those between ISMR and D20 based predictor ($D_p$) (Extended Data Figure 5c). As HC contains some influence of atmospheric 'noise' through surface temperature, the poorer correlations appear to be manifestations of this influence. The correlations between ISMR and $S_p$ are even smaller than those between ISMR and $H_p$ as the influence of the atmospheric 'noise' appears to be largest on the surface temperature.

# Forecast and Verification

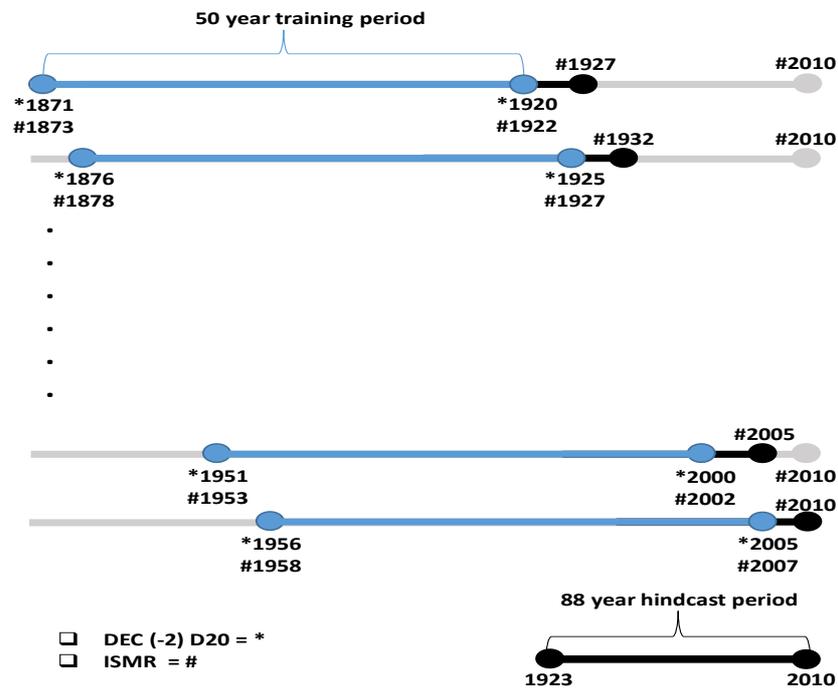

*Figure 6: A schematic diagram of the 18-month lead hindcast using DEC (-2) D20.*

As indicated by Delsol and Shukla (2009)[59] artificial skill may be built in the forecasts if the full data is involved in the predictor discovery. In order to keep a section of the data truly independent from development and training, while the potential for prediction is evaluated using the predictors discovered using the full data (1871-2010) (Figure 1), for hindcasts the predictors are constructed from a subset of D20 (1871-1977(-2), 1872-1978(-1), 1873-1979(0)) and ISMR (1873-1979) data so that the data between the period 1980 and 2010(ISMR) is not used either in development or training of the forecast model. The correlation maps similar to Extended Data Figure 1 are created based on ISMR (1873-2010) and D20 (or HC or SST) for the period 1871-2008(-2), 1872-2009(-1) and 1873-2010(0). The predictors are found by projecting the monthly D20 anomaly (or HC or SST anomaly) on these correlation maps for all lags as described above.

As the predictor-predictand relationship does have epochal variations, to minimize the influence of these epochal variations on the hindcasts, we employ a 50-year sliding window training and prediction of next 5 years up to 2010 similar to one used by Wang et al. (2015)[30]. The first training period is 1871-1920 and 1873-1922 with respect to DEC (-2) D20 and ISMR respectively and predictions for 1921-1925(ISMR), the next training period is 1876-1925 and 1878-1927 with respect to DEC (-2) D20 and ISMR respectively and predictions for 1926-1930 and so on. The hindcasts for the period between 1980 and 2010 (31 years) are based on training the model using DEC (-2) D20 and ISMR data for the period 1928-1977 and 1930-1979 respectively, making these hindcasts for 31-years partial independent hindcasts. Thus, we have 88-years of partially independent hindcasts. The potential skill of ISMR prediction arising from association with slow coupled oscillations in the tropical climate system is evident from the correlation between $D_p$ and ISMR (r = 0.84). The hindcasts indicate that the limit on the potential skill could be even higher (r = 0.86).

**Estimation of growth of errors:**

The predictability limit is governed by the growth rate of small errors and its saturation level. This could be easily estimated from a coupled model by conducting identical twin experiments[60] and has been shown that the growth rate of errors in the tropical coupled system is governed by two time scales. Here, we attempt to estimate growth of initial errors of the tropical coupled system from observations. As shown by Goswami and Xavier (2003)[62], it is possible to do so by estimating the event-to-event variability and diversity of their evolution. However, unlike coupled model experiments, there is no possibility of controlling the amplitude of 'initial error' in this case. Following Goswami and Xavier (2003)[62], we estimate the growth of error in the following way: first we identified all the peak ENSO events defined by PC1 of NDJ D20 EOF from 1871 to 2010 being larger than 1-s.d. This way 20 such events are identified. Taking any initial condition with respect to the peak (say the peak itself Dec (-1)), the evolution of D20 anomalies up to 24 months ahead for all events, being their trajectory of evolution, are extracted. A plot of standard deviation as a function of lead-time would give an estimate of event-to-event variability and a measure of divergence of trajectories or growth of errors. This is done at each grid point. The zonal means of growth of errors are shown in Figure 4c.

**Extended Figures and Table:**

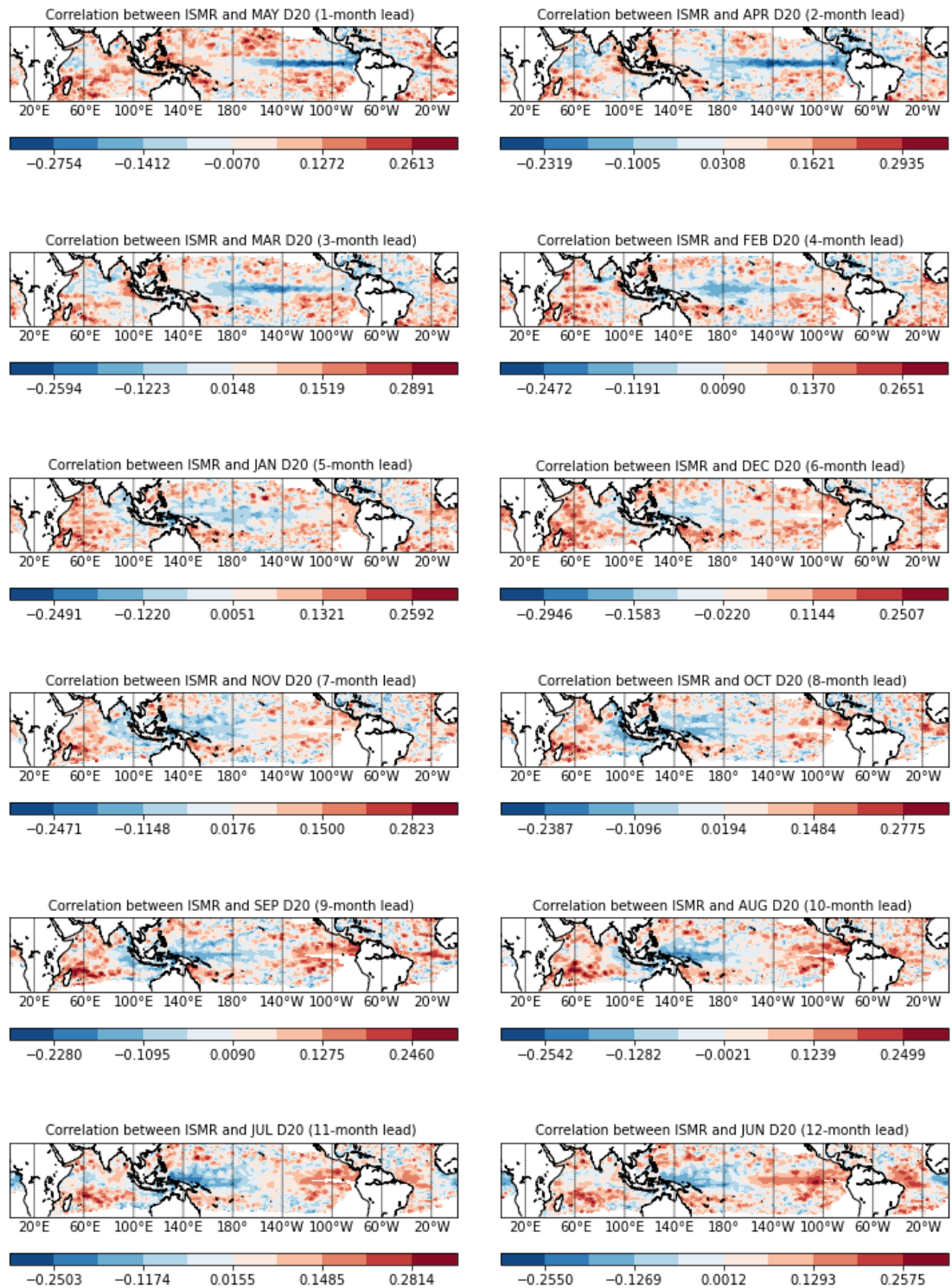

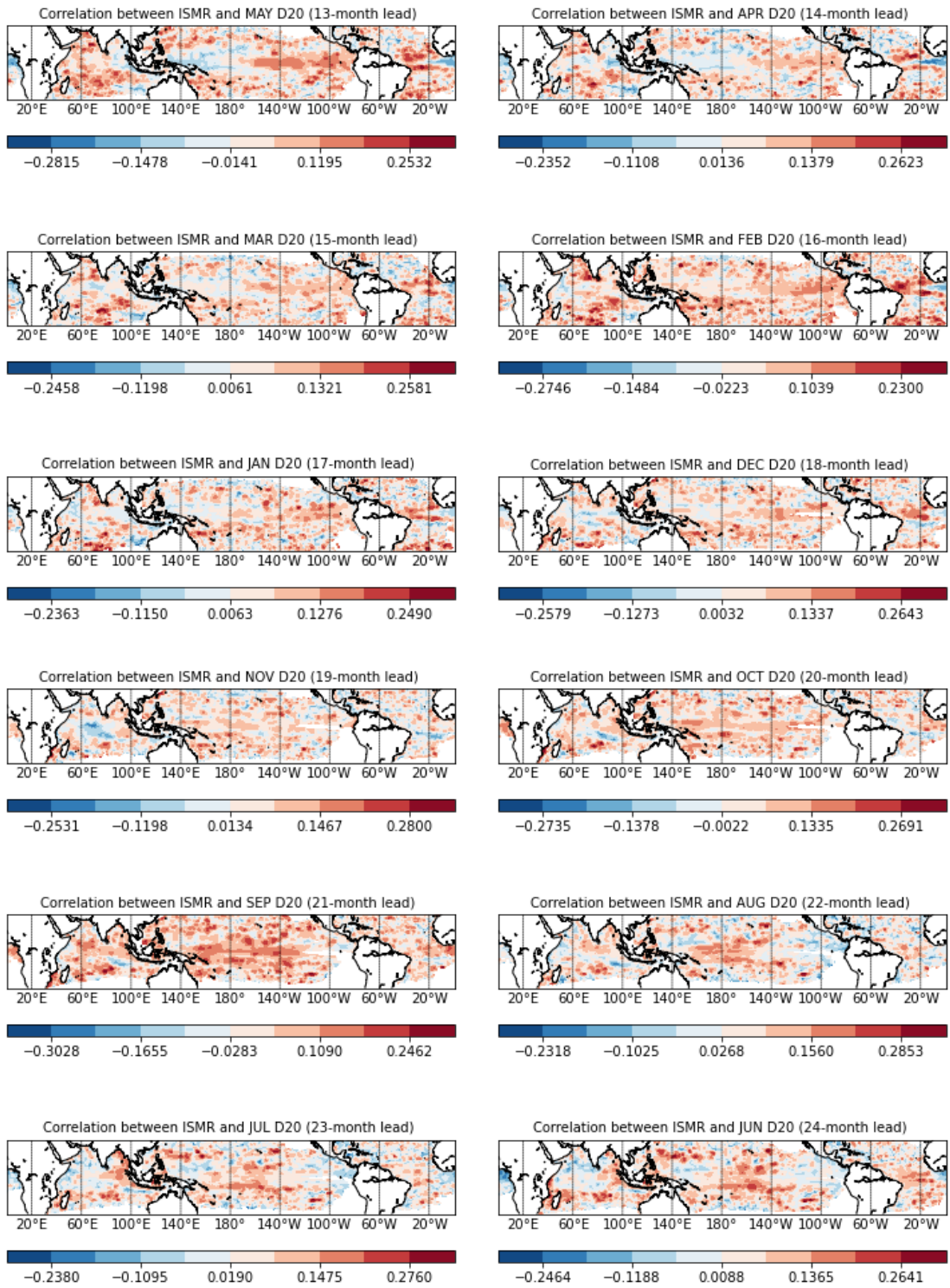

*Extended Data Figure 1: With JJAS as the ISMR year (0) season, May (0) is lag = 1, April (0) is lag=2 months and so on. Lags and corresponding months with respect to ISMR are shown in 24 Maps of correlations between ISMR and monthly D20 anomaly at lags up to 24 months.*

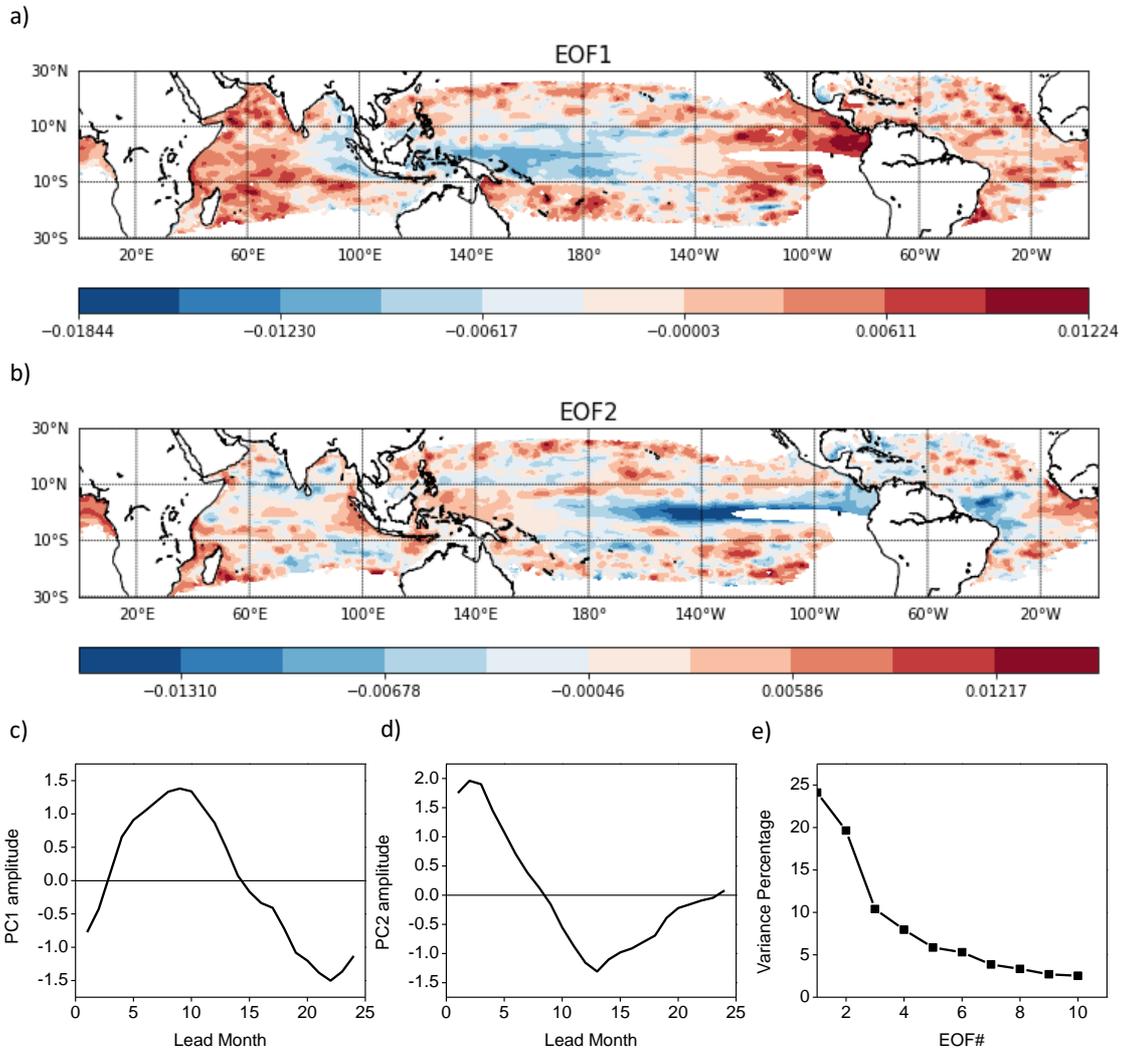

Extended Data Figure 2: Two dominant EOFs of CorD for the 24 lags. (a) EOF1, (b) EOF2, (c) PC1, (d) PC2, (e) Variance explained by the leading EOFs.

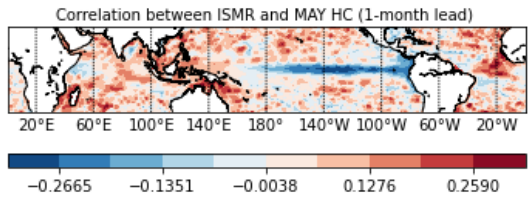
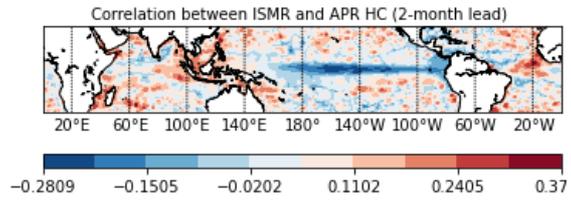
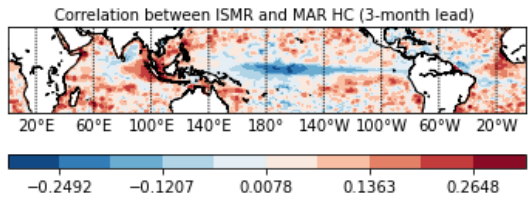
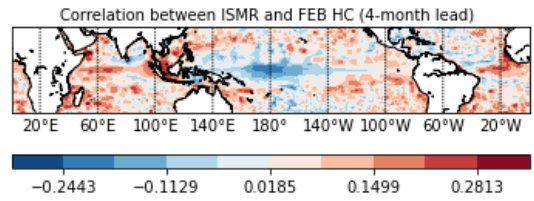
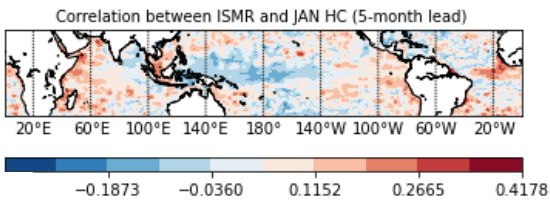
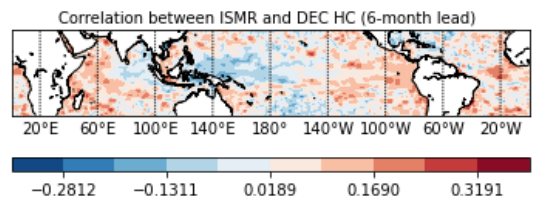
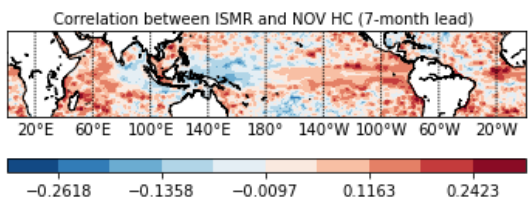
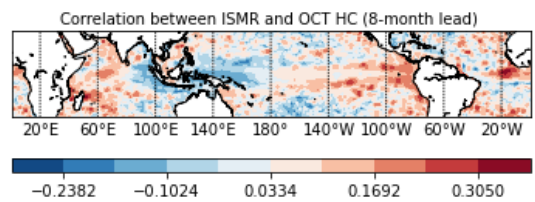
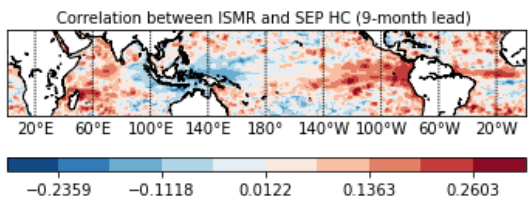
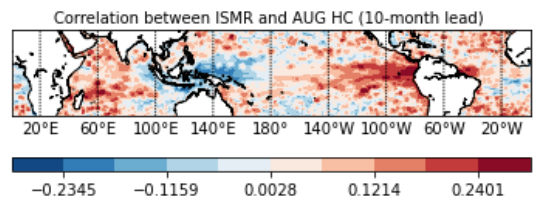
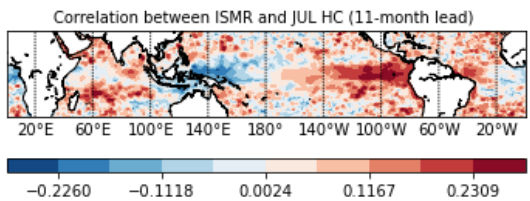
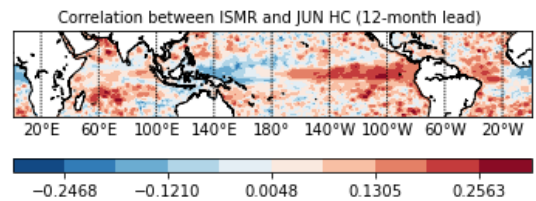

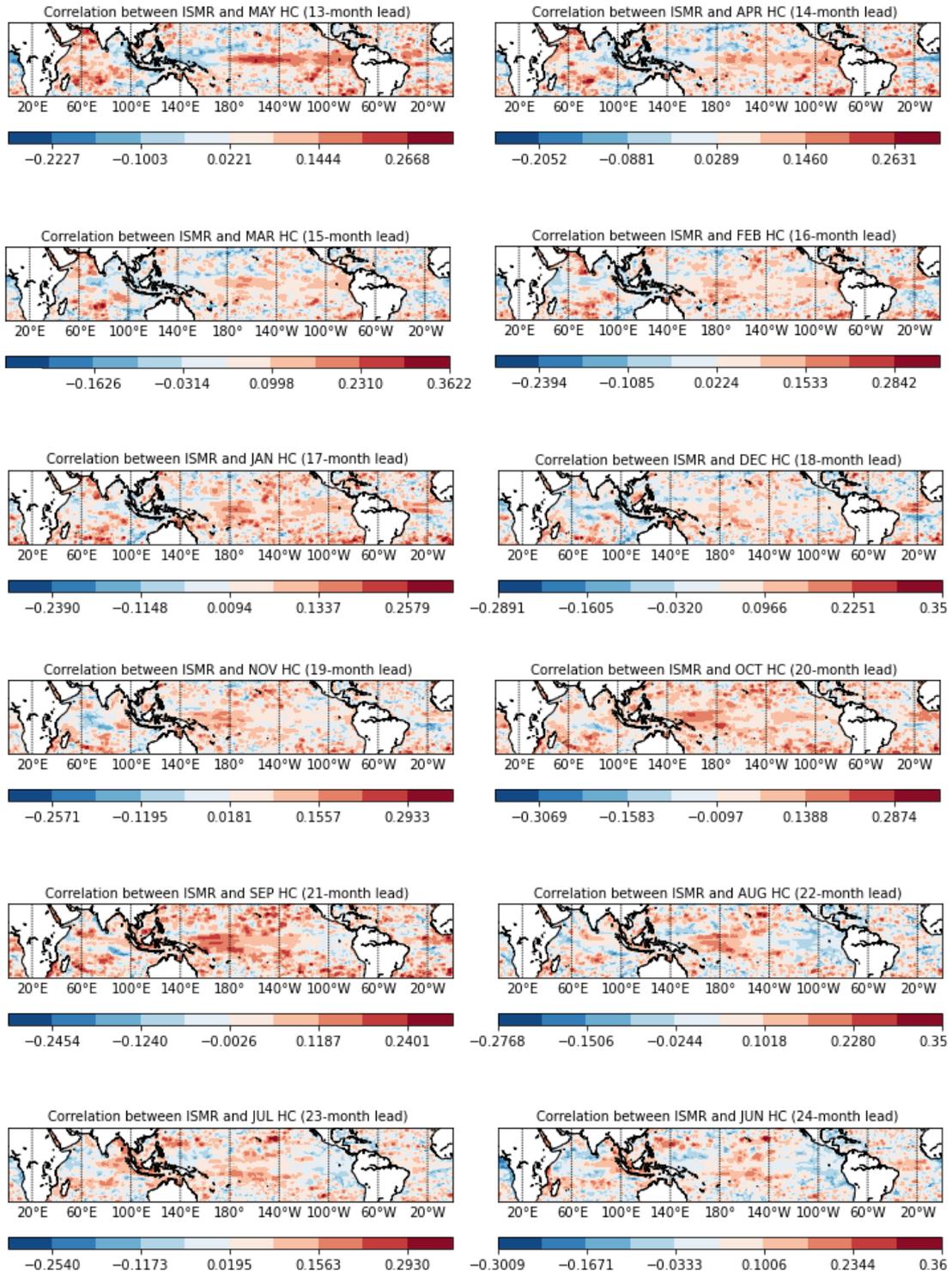

*Extended Data Figure 3: Similar to Extended Data Figure 1 but for correlations between ISMR and monthly HC anomaly at lags up to 24 months.*

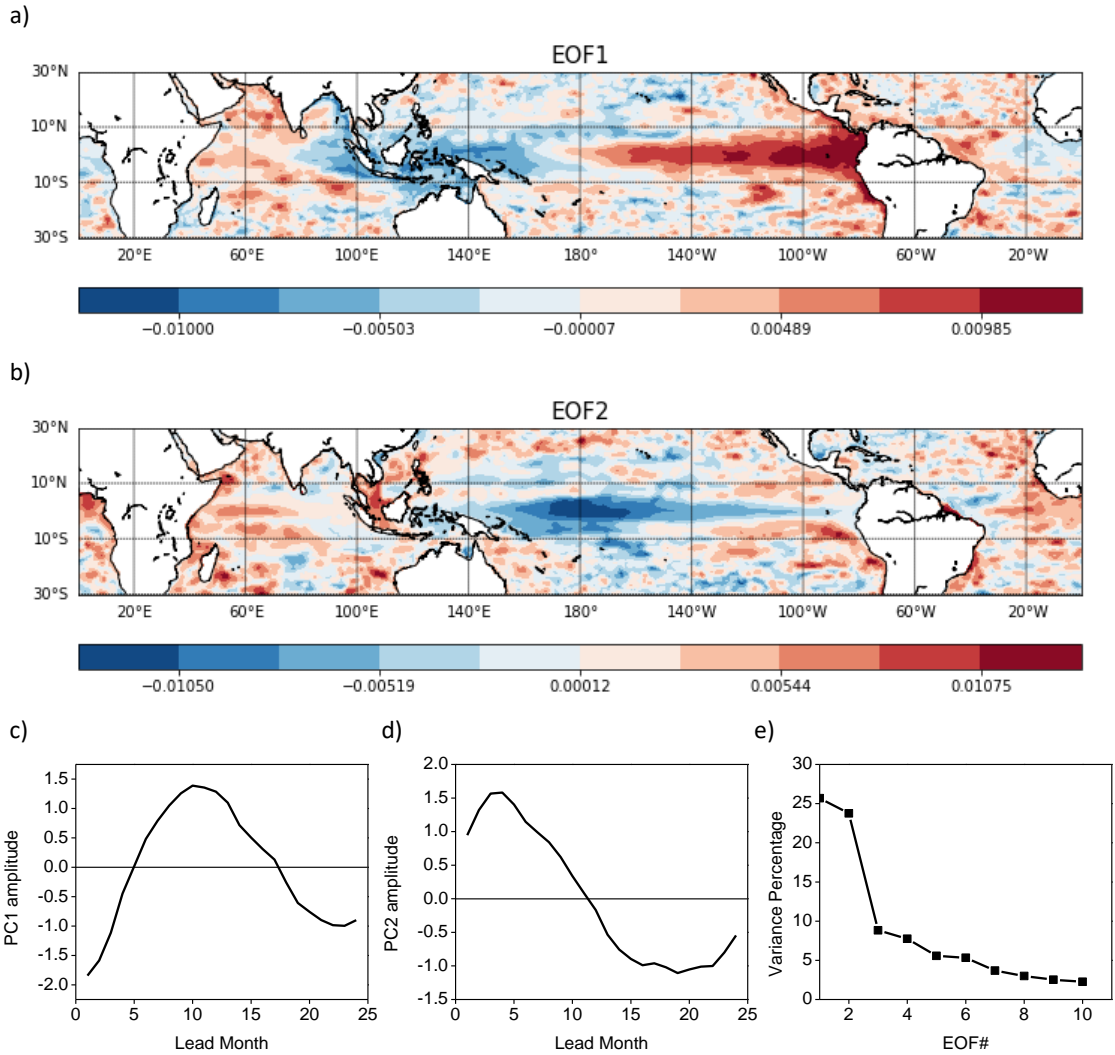

*Extended Data Figure 4: Similar to Extended Data Figure 2 but for CorH.*

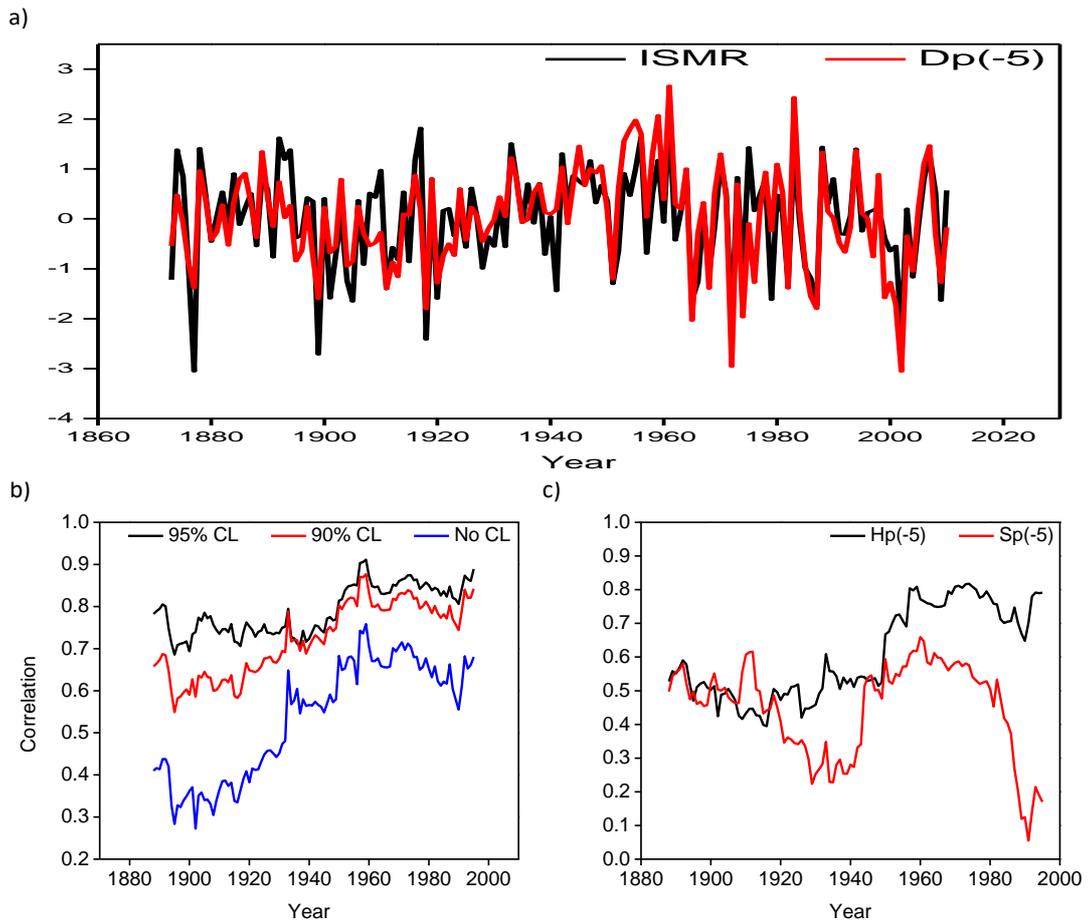

*Extended Data Figure 5: (a) Time series of normalized ISMR anomaly and $D_p(-5)$ i.e the D20 monthly anomaly based predictor with 5-month lead corresponding to Jan(0) for the period 1871 to 2010. Simultaneous correlation between the two indicated. (b) 31-year moving correlation between ISMR and $D_p(-5)$ constructed using projections on the full tropical correlation pattern (no CL), on pattern exceeding 90% CL and on pattern exceeding 95% CL. (c) 31- year moving correlations between ISMR and $H_p(-5)$ and with $S_p(-5)$.*

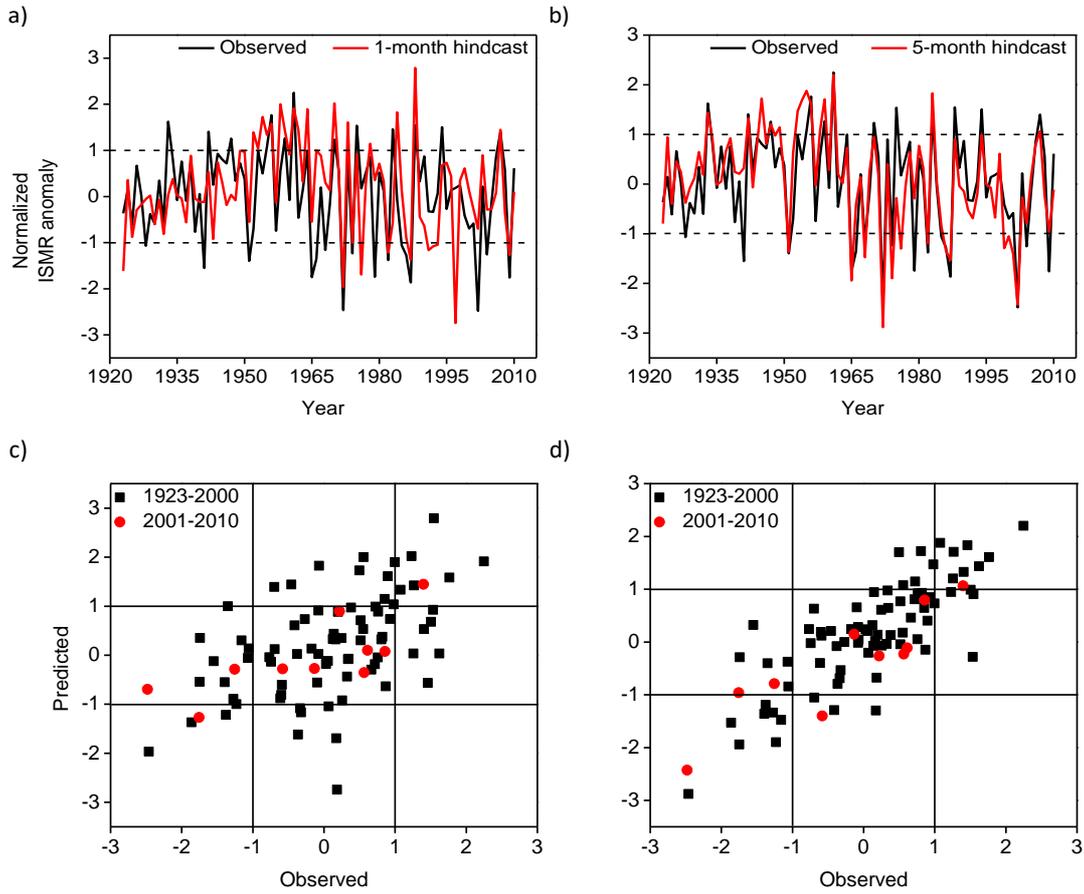

*Extended Data Figure 6: Verification of 1-month lead and 5-month hindcasts. (a) & (b) Time series of predicted and observed normalized ISMR anomalies for 1-month and 5-month lead respectively. (c) & (d) Scatter plot of predicted and observed normalized ISMR anomalies for 1-month and 5-month lead respectively.*

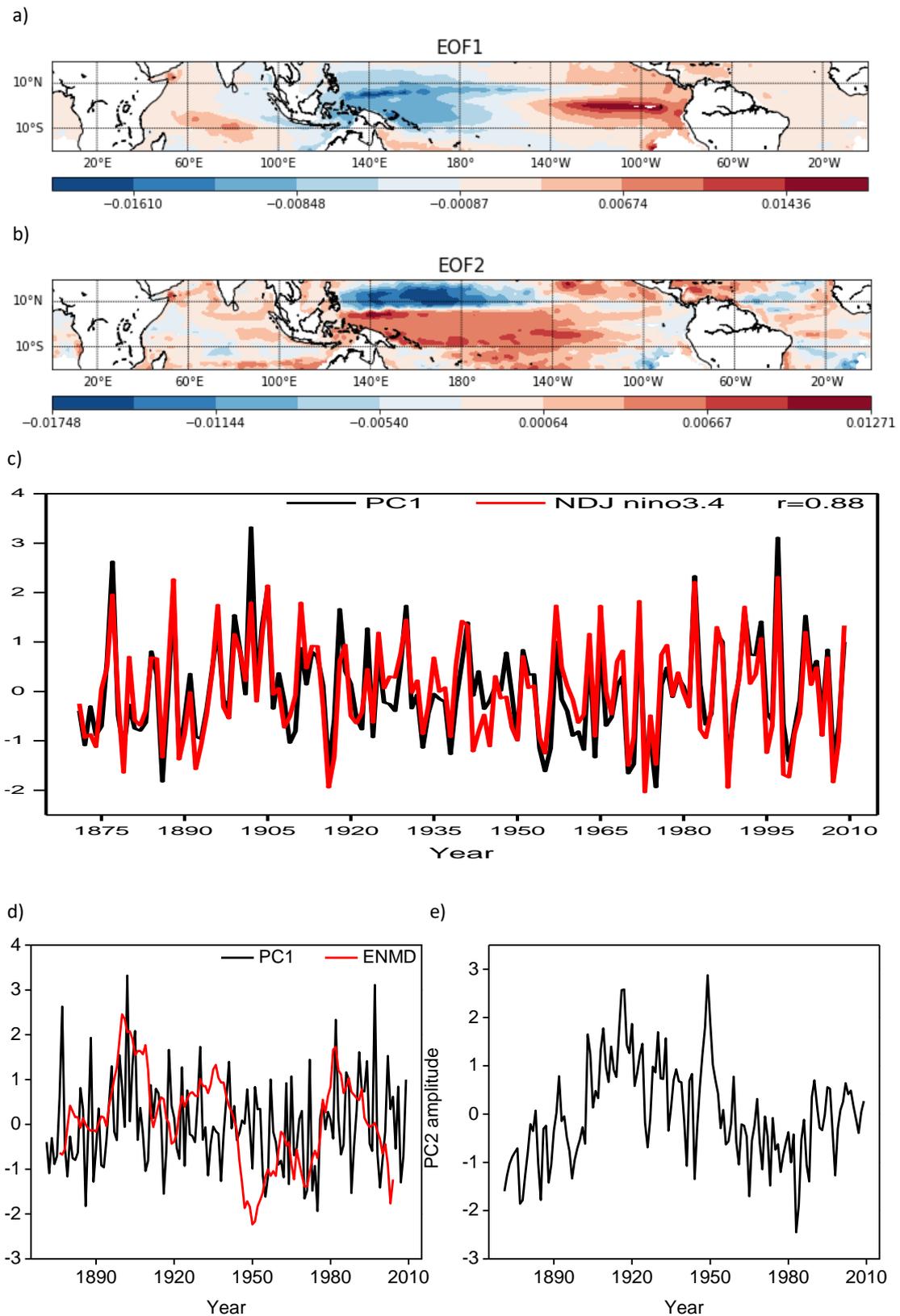

*Extended Data Figure 7: Empirical Orthogonal Function (EOF) analysis of NDJ D20 anomaly over the tropical domain (20⁰S-20⁰N) between 1871 and 2010. (a) EOF1, (b) EOF2, (c) PC1 together with NDJ Nino3.4 SST anomaly normalized by their own interannual standard deviation. The correlation between PC1 and Nino3.4 is r = 0.88. (d) Normalized PC1 and normalized ENMD. (e) The normalized PC2.*

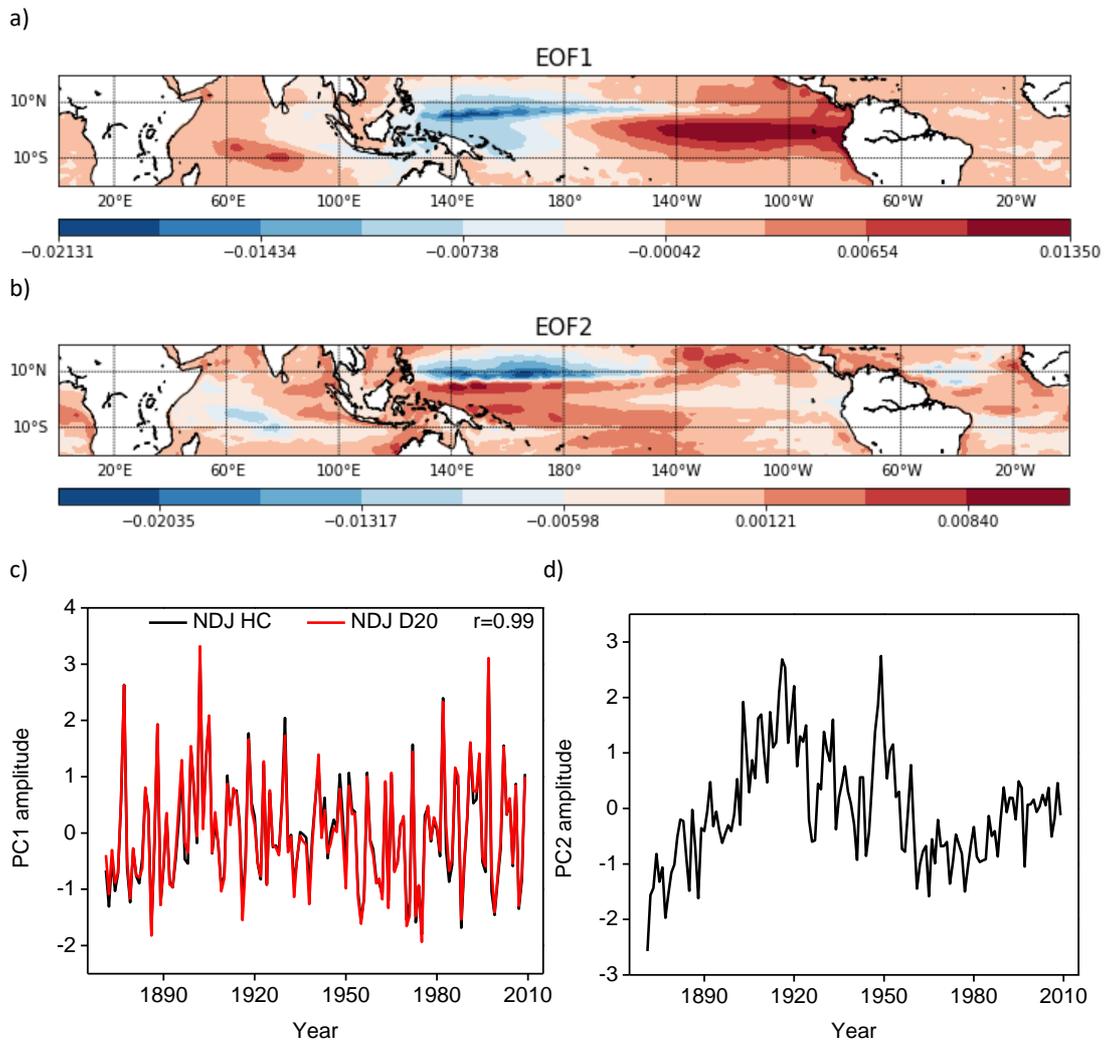

*Extended Data Figure 8: Empirical Orthogonal Function (EOF) analysis of NDJ HC anomaly over the tropical domain (20⁰S-20⁰N) between 1871 and 2010. (a) EOF1, (b) EOF2, (c) PC1 together with PC1 of NDJ D20 anomaly (Extended Data Figure 7c) normalized by their own interannual standard deviation. The correlation between the two is r = 0.99. (d) Normalized PC2*

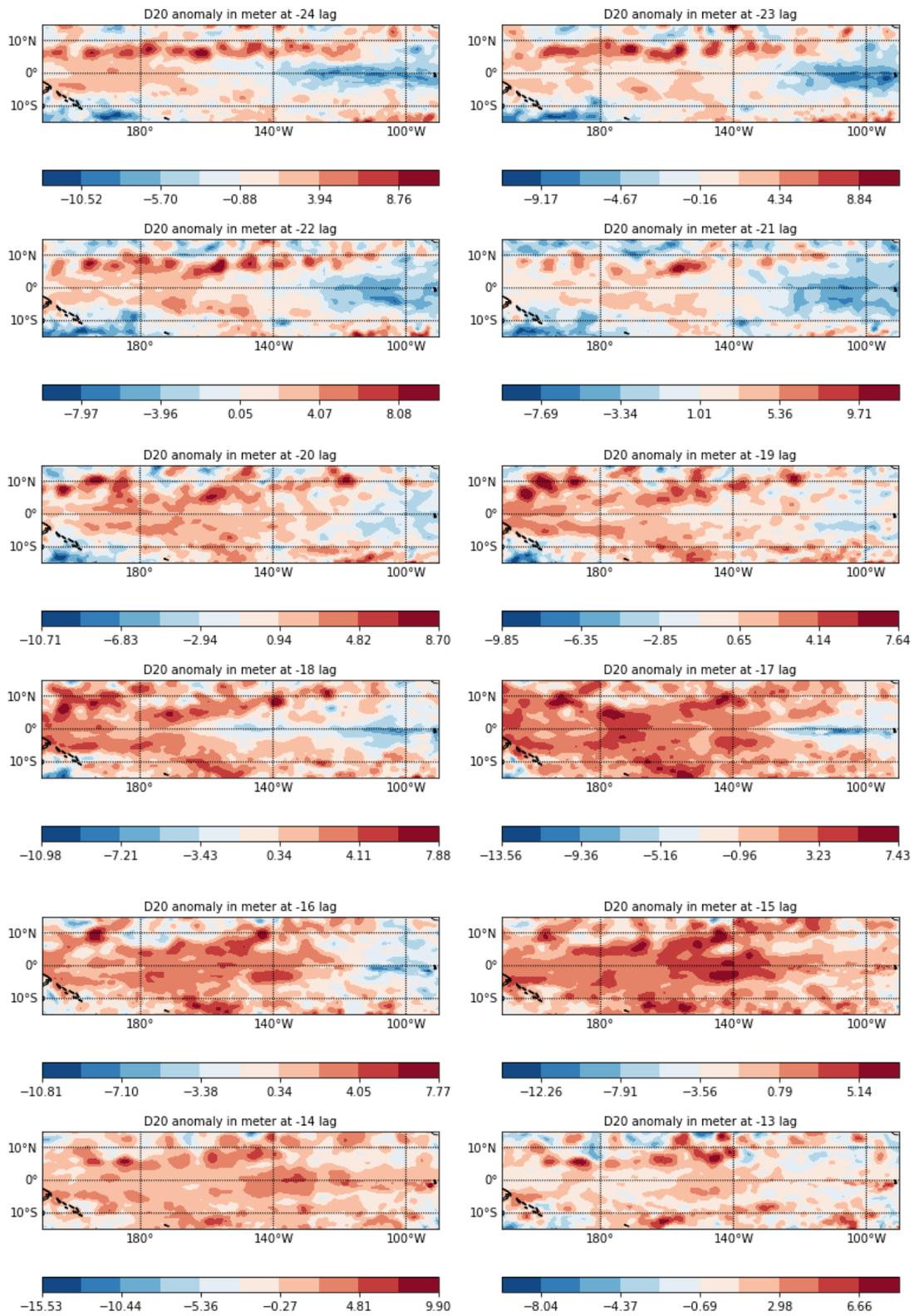

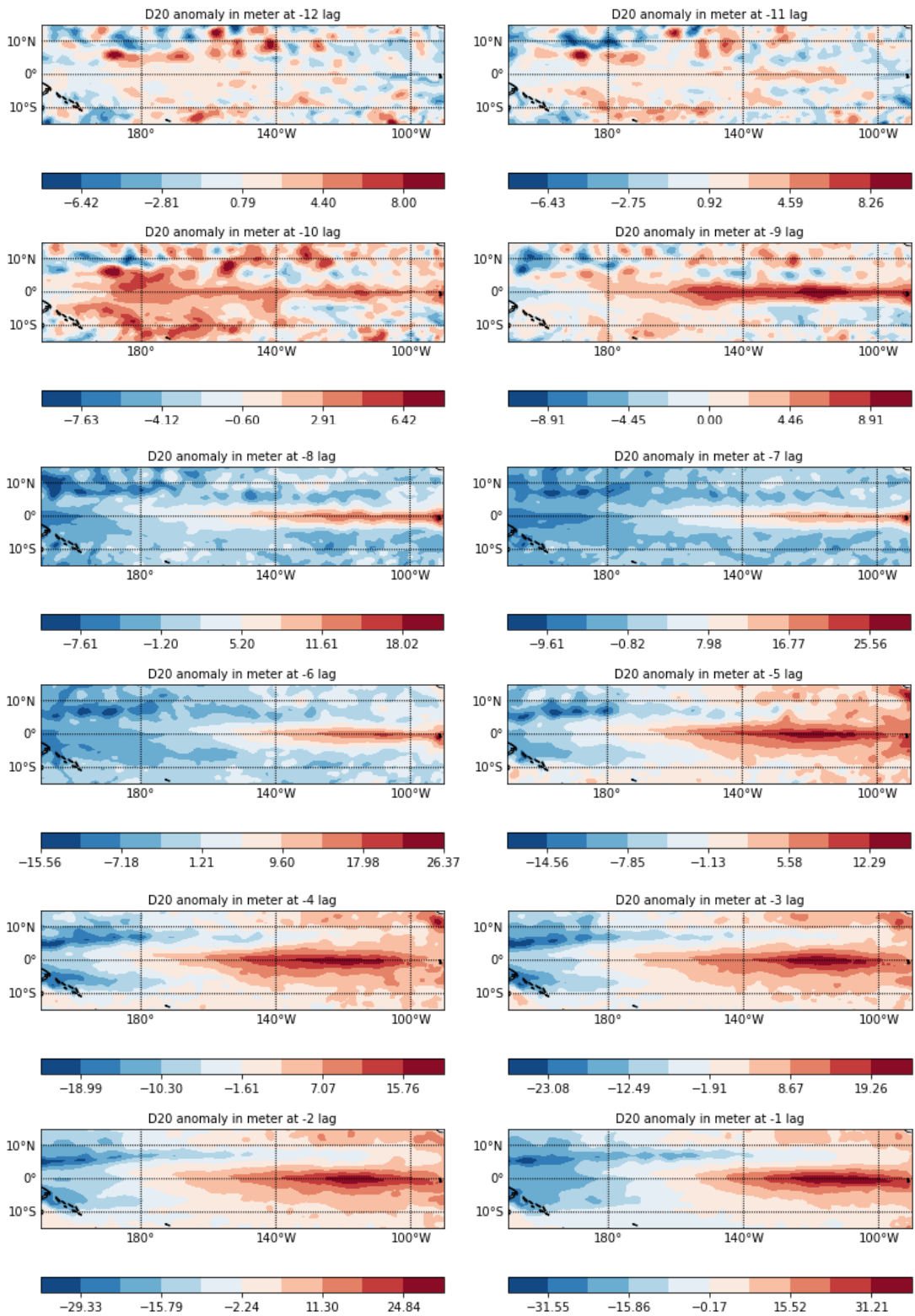

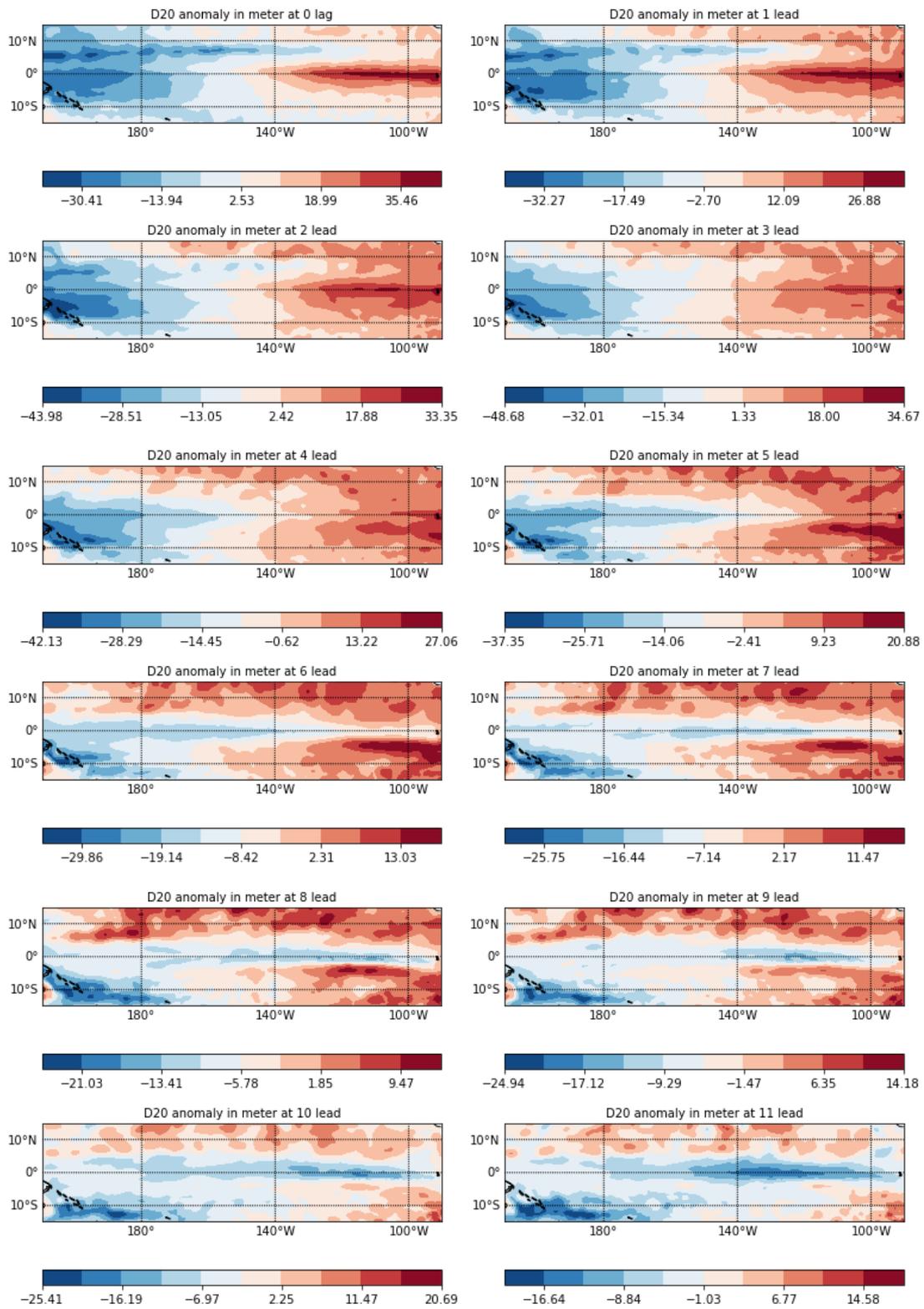

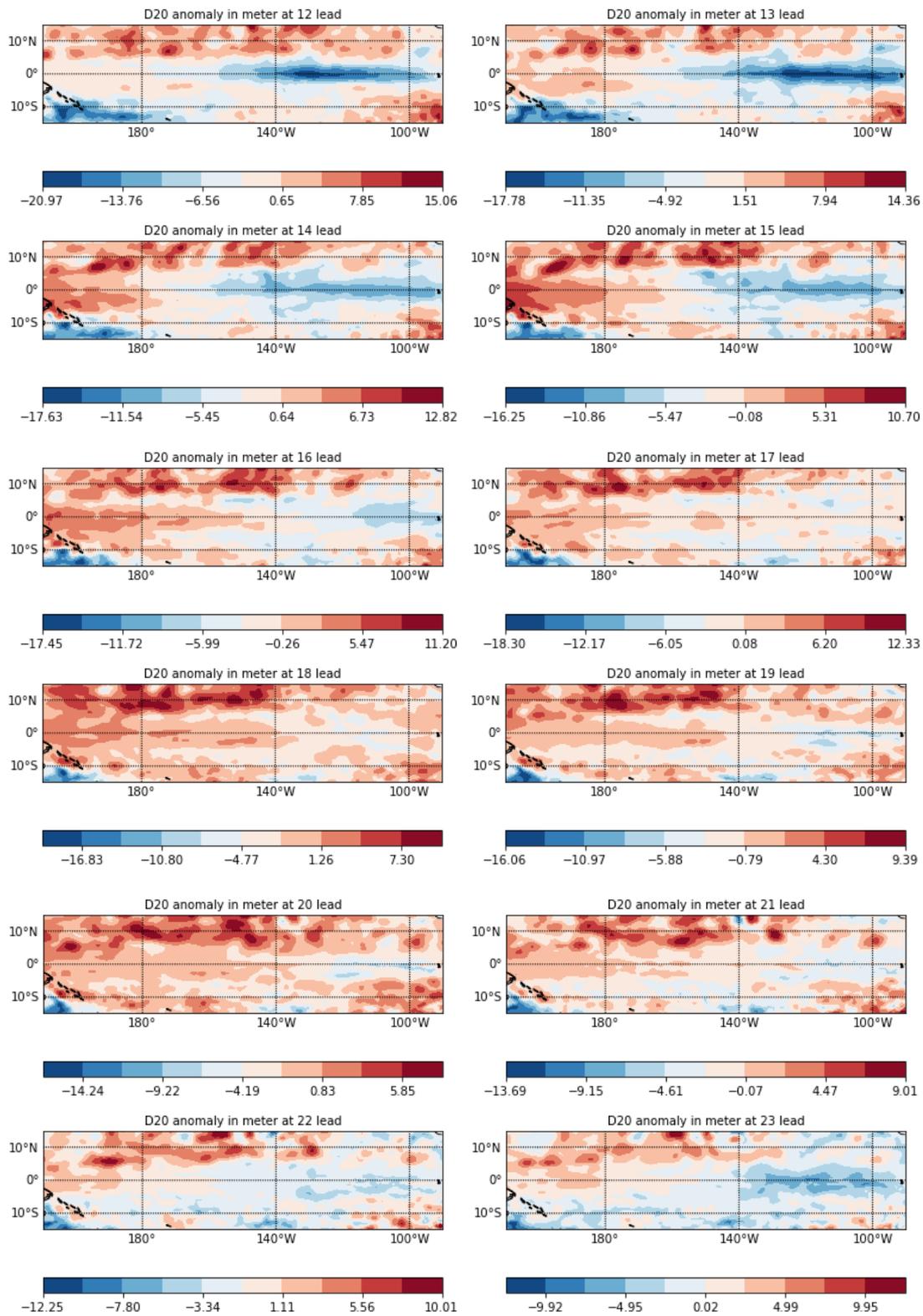

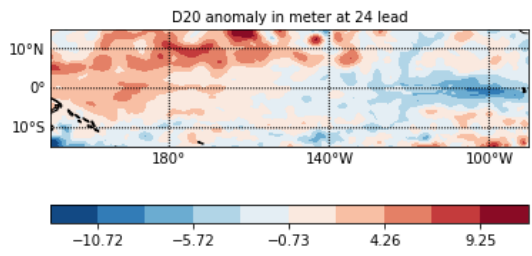

Extended Data Figure 9: Lead-lag composite of monthly D20 anomaly (m) with respect to peak ENSO events identified from PC1 of NDJ D20 anomaly (Extended Data Figure 7c from 24 month lags to 24 month leads around Dec (0).

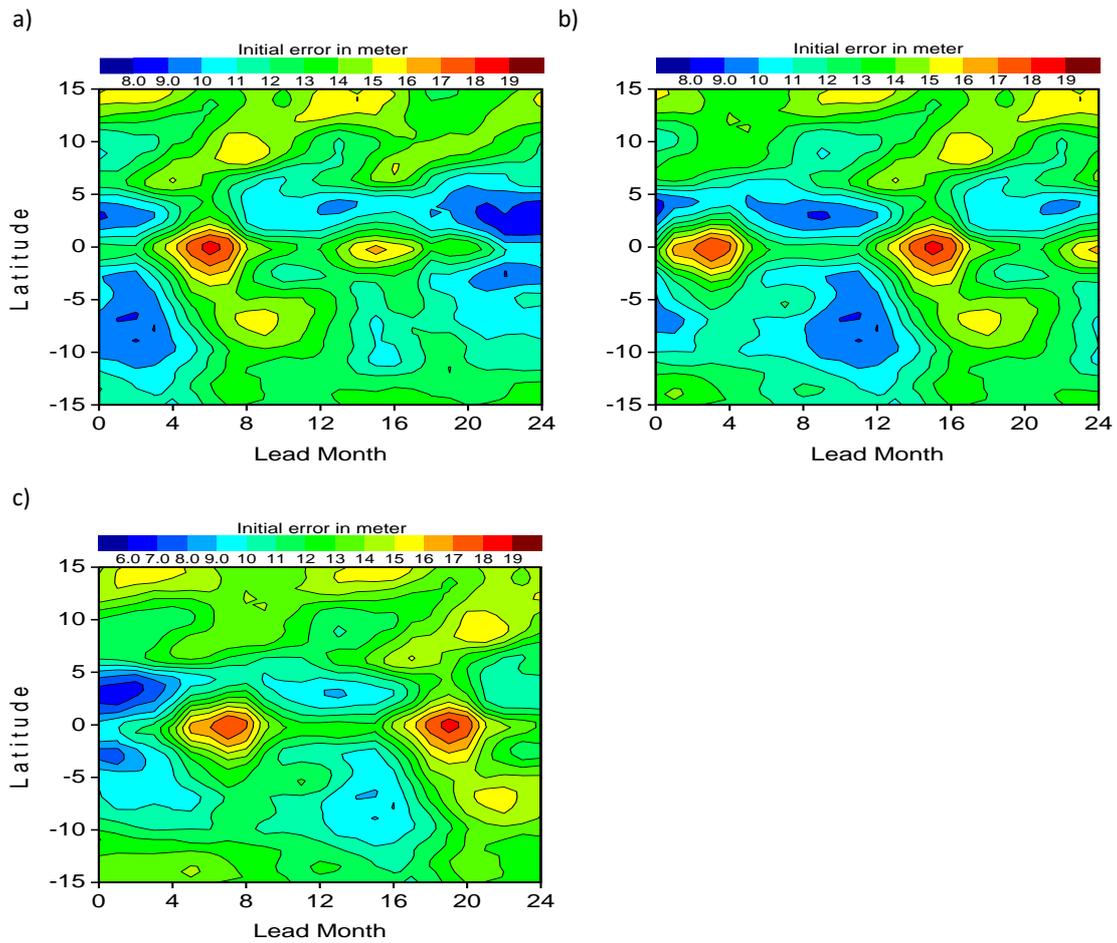

*Extended Data Figure 10: (a) Zonal mean over Pacific basin of estimated growth of initial error (standard deviation of divergence of trajectories) up to a lead of 24-months for an initial condition corresponding to 3-month before the peak of ENSO event (Sept (-1)) as a function of latitudes. (b) Same as (a) but for initial conditions corresponding to 12-month before the peak ENSO event and (c) Same as (a) but for initial conditions corresponding to 16-month before the peak ENSO event.*

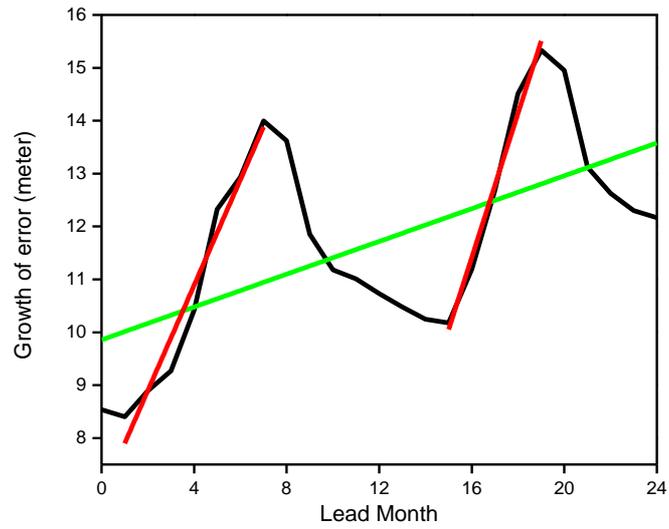

*Extended Data Figure 11: Zonal mean and latitudinal mean between 7⁰S and 5⁰N of estimated growth of errors for an initial condition corresponding to peak of El Nino event for an initial condition corresponding to 16-months before the peak of El Nino event (Aug (-2)). A fast rate of growth (red) and slow rate of growth (green) of errors are indicated.*

| Lead Month | Confidence level | Correlation | RMSE (mm) | Standard deviation (mm) |
|---|---|---|---|---|
| 1 (MAY(0)) | 95% | 0.52 | 66.32 | 45.05 |
| 1 (MAY(0)) | 90% | 0.58 | 63.17 | 49.06 |
| 1 (MAY(0)) | No | 0.66 | 58.20 | 53.53 |
| 5 (JAN(0)) | 95% | 0.81 | 47.20 | 75.08 |
| 5 (JAN(0)) | 90% | 0.78 | 49.42 | 69.46 |
| 5 (JAN(0)) | No | 0.62 | 61.64 | 54.37 |
| 18 (DEC(-2)) | 95% | 0.86 | 41.94 | 80.31 |
| 18 (DEC(-2)) | 90% | 0.86 | 42.39 | 81.84 |
| 18 (DEC(-2)) | No | 0.84 | 45.54 | 84.03 |

*Extended Data Table 1: Correlations, RMSE and standard deviation of hindcasts for three leads, 1-month lead (May(0)), 5-month lead (Jan (0)) and 18-month lead (Dec(-2) for predictors based on projections over the full correlation pattern (No CL) and on correlation patterns exceeding 90% and 95% CL.*